\documentstyle{siam}

\def\DD{\mbox{$DD$}}
\def\DDB{\mbox{$DD_{\cal B}$}}
\def\Tr{\mbox{$TT$}}
\def\mdp{\mbox{$PTO$}}
\def\mdg{\mbox{$GTO$}}
\def\mpd{\mbox{$PTO$}}
\def\mgd{\mbox{$GTO$}}

\def\maxactivity{n+3t}
\newcommand{\overj}{\bar{\mbox{{\em \j}}}}
\newcommand{\ac}{{\it ac}}
\def\overk{\bar{k}}
\newcommand{\overi}{\bar{\mbox{{\em \i}}}}
\def\overl{\bar{l}}
\def\A{\mbox{$\cal A$}}
\def\B{\mbox{$\cal B$}}
\def\C{\mbox{$\cal C$}}
\def\D{\mbox{$\cal D$}}

\def\mod{{\rm mod~}}
\def\goahead{\mbox{\bf go ahead}}
\def\gets{:=}
\def\grade{\mbox{\em grade}}
 
\def\curr{\hbox{{\it current round\/}}}
\def\false{\hbox{{\sc false\/}}}
\def\found{\hbox{{\sc done\/}}}
\def\point{\hbox{{\sc point\/}}}
\def\pointer{\hbox{{\sc point\/}}}
\def\round{\hbox{{\sc round\/}}}
\def\succ{\hbox{{\it successor\/}}}
\def\true{\hbox{{\sc true\/}}}
\def\onlytwice{\hbox{{\sc round\/}}}

\def\done{\hbox{{\sc done\/}}}

\newcommand{\eprf}{\bbox\vspace{0.1in}}
\newcommand{\bbox}{\vrule height7pt width4pt depth1pt}
\newcommand{\QED}{\eprf}

\newtheorem{claim}{Claim}[section]
\newcommand{\remove}[1]{}
\newcommand{\suppress}[1]{}
\long\def\notefromO#1{}

\newcommand{\STOC}[1] {Proc. #1 ACM Symposium on Theory of
Computing}
\newcommand{\PODC}[1] {Proc. #1 ACM Symposium on
Principles of Distributed Computing}
\newcommand{\FOCS}[1] {Proc. #1 IEEE Symposium on Foundations
of Computer Science}
\newcommand{\JACM}[1] {{\em Journal of the ACM}}

\begin{document}
\title{PERFORMING WORK EFFICIENTLY IN THE PRESENCE OF
FAULTS\thanks{A preliminary version of this work appeared in
\PODC{11th}, 1992.  This version will appear in {\em SIAM Journal on
Computing}. 
During this research, Orli Waarts  was
at Stanford University and supported by U.S. Army Research Office
Grant DAAL-03-91-G-0102, NSF grant CCR-8814921, ONR contract
N00014-88-K-0166, and an IBM fellowship.}
}
\author{
Cynthia Dwork\thanks{IBM Research Division,
Almaden Research Center, K53-B2,
650 Harry Road,
San Jose, CA 95120-6099. {\tt dwork@almaden.ibm.com}}
\and
Joseph Y. Halpern\thanks{IBM Research Division,
Almaden Research Center, K53-B2,
650 Harry Road,
San Jose, CA 95120-6099.
{\tt halpern@almaden.ibm.com}}
\and
Orli Waarts\thanks{Computer Science Division,
U.C. Berkeley, Berkeley, CA 94720. {\tt
waarts@cs.berkeley.edu}}
}
 
\maketitle
\begin{abstract}
We\notefromO{Added footnote to title. Changed affiliation.}
consider a system of
$t$ synchronous processes that communicate only by sending
messages to one another, and that together
must perform $n$ independent units of work.
Processes may fail by crashing;
we want to guarantee that
in every execution of the protocol
in which at least one process survives, all $n$ units of
work will be performed.
We consider three parameters: the number of messages sent,
the total number of units of work performed (including
multiplicities), and time.
We present three protocols for solving the problem.
All three are
work-optimal, doing $O(n+t)$ work.
The first
has moderate costs in the remaining two
parameters,
sending $O(t\sqrt{t})$ messages, and
taking $O(n+t)$ time.
This protocol can be easily modified to
run in any completely
asynchronous system equipped with a failure detection
mechanism.
The second
sends only $O(t\log{t})$ messages, but
its running time is large ($O(t^2(n+t) 2^{n+t})$).
The third is essentially time-optimal in the (usual) case
in which there are no failures, and its
time complexity degrades gracefully as the number of
failures increases.
\end{abstract}

\begin{keywords}
fault-tolerance, work, Byzantine agreement, load balancing, distributed systems
\end{keywords}
 
\section{Introduction}
A fundamental issue in distributed computing is
fault-tolerance:
guaranteeing
that work is performed, despite the presence of failures.
For example,
in controlling a nuclear reactor
it may be crucial for a set of valves to be closed before
fuel is added.
Thus,
the procedure for
verifying that the valves are
closed must be highly fault-tolerant.
If processes never fail then the {\em work} of
checking that the valves are closed
could be distributed according to some load-balancing technique.
Since processes
may fail,
we would like an algorithm that guarantees
that the work will be performed
as long as at least
one process survives.
Such an algorithm could be particularly useful in a local area network,
where jobs might be distributed among idle workstations.  (The idea of
running computations on idle nodes is an old one, going back at least to
\cite{SH}.  See \cite{GK} for one implementation of this
approach, and further references.)  In this
case a ``failure'' might correspond to a user reclaiming her machine.

The notion of work in this paper is very broad, but is restricted
to ``idempotent'' operations, that is,
operations that can be repeated without harm.
This is because if a process performs
a unit of work and fails before telling a second process of
its achievement, then the second process has no choice but
to repeat the given unit of work.
Examples include verifying a step in a formal proof,
evaluating a boolean formula at a particular assignment
to the variables, sensing the status of a valve,
closing a valve,
sending a message, say, to a process outside of
the given system, or
reading records in a distributed database.
 
Formally, we assume that we have a synchronous system of $t$
processes that are subject to crash failures, that want to
perform $n$ independent units of work. (For now, we assume that initially there is common
knowledge among the
$t$ processes about the $n$ units of work to be performed.  We
return to this point later.) In one time unit a process
can compute locally and perform one unit of work and one round of
communication (sending and receiving messages).\notefromO{Added
this sentence.} Given that performing a unit of work
can be repeated without harm, a trivial solution is obtained by
having each process perform
every unit of work.
In our original example, this would mean that every process
checks that every valve is closed.
This solution requires no messages, but in the worst case
performs $tn$ units of work
and runs in $n$ rounds.
(Here the worst case is when no process fails.)
 
Another straightforward solution can be obtained by having only
one process performing the work at any time, and checkpointing
to each process after completing every unit of work.
In this solution, at most $n+t-1$ units of
work are ever performed,
but the number of messages sent
is almost $tn$ in the worst case.
 
In both these solutions the total amount of
{\em effort}, defined as work plus messages, is $O(tn)$.
If the actual cost of performing a unit of work is comparable
to the cost of sending a message,
then neither solution is
appealing.
In this paper we focus on
solutions which are work-optimal, up to a constant factor,
while keeping the total effort reasonable.
Clearly, since a process can fail immediately after
performing a unit of work, before reporting that
unit to any other process, a work-optimal
solution performs $n+t-1$ units of work in the worst case.
Thus, we are interested in solutions that perform $O(n+t)$ work.
 
Let $n' = \max(n,t)$.
Our first result is an algorithm whose total effort
is at most $3n'+9t\sqrt{t}$.  In fact, in the worst case
the amount of work performed is at most $3n'$ and the
number of messages is at most $9t\sqrt{t}$, so the form
of the bound explains the costs exactly.
We then optimize this algorithm to achieve
running time of $O(n+t)$ rounds. Note
that any solution requires
$n$ rounds in the worst case, since if $t-1$ processes are initially
faulty then the remaining process must perform all
$n$ units of work.
In this algorithm the synchrony is used only to detect failures,
as usual by detecting the absence of an expected message.
Thus, it can be easily modified
to work in a completely asynchronous system
equipped with a failure detection mechanism.

We then prove that the above algorithm is
not message-optimal (among work-optimal algorithms),
by constructing
a technically challenging work-optimal algorithm
that requires only $O(t\log t)$ messages in the worst case.
Since $O(n + t)$ is a lower bound on work,
and hence on effort,
the $O(n+t \log t)$ effort of this algorithm is
nearly optimal.
The improved message complexity is obtained
by a more
aggressive use of synchrony.
In particular, the absence of a message in this algorithm
has two possible meanings: either the potential sender failed or
it has insufficient ``information''
(generally about the history of the execution),
and therefore
has chosen not to send a message.
Due to this use of synchrony, unlike the first algorithm,
this low-effort
algorithm will not run in the asynchronous model with
failure detection.
In addition,
the efficiency comes at a price in
terms of
time: in the worst case,
the algorithm requires $O(t^2(n+t) 2^{n+t})$ rounds.
 
The first two algorithms are very sequential:
at all times work is performed by a single active process
who uses some checkpointing strategy to inform other
processes about the completed work.
This forces the algorithms to take at least $n$
steps, even in a failure-free run.
To reduce the time we need to increase parallelism.
However, intuitively, increasing parallelism while
simultaneously minimizing time and remaining work-optimal
may
increase communication costs, since processes must
quickly tell
each other about completed work.
The third algorithm does exactly this in a fairly
straightforward way,
paying
a price in messages
in order to decrease best-case time.
It is designed to perform
time-optimally
in the absence of failures,
and to have its time complexity
degrade gracefully with additional faults.
In particular, it takes
$n/t+2$ rounds in the failure-free case,
and its message cost is $2t^2$; its worst-case message cost
is $O(f t^2)$,
where $f$ is the actual number of failures in the
execution.
 
There are a number of assumptions in our model
that are arguably not
realistic.  For one, we assume that the $n$ units of work are
identical, or, at least, that they all take the same amount of
time to perform.  In addition, we assume that the total workload
is static, and is common knowledge
at the beginning of the algorithm.  It is not
too hard to modify our last algorithm to deal with a more realistic
scenario, where work is continually coming in to different sites of
the system, and is not initially common knowledge.
We remark that a patent has been filed by IBM for such a modified
algorithm.
 
One application of our algorithms is to
Byzantine agreement.
The idea is that the general tries to
inform $t$ processes, and then each
of these
$t$ processes performs the ``work'' of ensuring that all
processes are informed.
In particular, our second algorithm,
called Protocol $\B$, gives a Byzantine
agreement algorithm for the crash fault model that requires $O(n + t\sqrt{t})$
messages and $O(n)$ time, where $n$ is the number of processes in
the system and $t$ is a bound on the number of failures,
while our third algorithm gives a
Byzantine agreement algorithm that uses $O(n + t\log{t})$ messages
and exponential time.
The best  result prior to ours was a nonconstructive
algorithm due to Bracha that requires $O(n+t \sqrt{t})$
messages \cite{Bracha}.
Galil, Mayer, and Yung \cite{GMY95} have recently
obtained an algorithm that uses
only a linear number of messages.
However, the algorithm is incomparable to
the agreement algorithm obtained using Protocol~$\B$ because
it requires a superlinear number of rounds.

Using the observation that our solutions to the work problem
yield solutions to Byzantine agreement,
we can now return
to the assumption that initially
there is common knowledge about the
work to be performed.
Specifically, if even one process knows about this
work, then it can act as a general, run Byzantine
agreement on the pool of work using one of the
three algorithms, and then the actual
work is performed by running the same algorithm a second
time on the real work.
If $n$, the amount of actual work, is $\Omega(t)$,
then the overall cost at most doubles when
the work is not initially common knowledge.
 
\subsection{Related Work}
 
The\notefromO{Made a subsection.} idea of doing work in the
presence of failures, in a different context, has appeared
elsewhere. First, Bridgland and Watro \cite{BW} considered a
system of $t$ asynchronous processes that together must perform
$n$ independent units of work. The processes may fail by crashing
and each process can perform at most one unit of work. They
provide tight bounds on the number of crash failures that can be
tolerated by any solution to the problem.
 
Clearly, our problem assumes a very different model than the one of
\cite{BW}. Furthermore, they want a protocol that guarantees that the work will
be performed in every execution of the protocol, while we want only a protocol
that guarantees that the work will be performed in executions in which at least
one process survives.
Consequently, their problem is not always solvable
and their focus is on finding conditions under which it is solvable.
Our problem is always solvable;
our focus is on finding efficient solutions.
 
Another similar but not identical problem was considered by
Kanellakis and Shvartsman. In a seminal
paper~\cite{KS} they consider the {\em Write-All}
problem, in which $n$ processes cooperate to set all $n$ entries of an
$n$-element array to the value~1. They provide an
efficient solution that tolerates up to $n-1$ faults, and
show how to use it to derive robust versions of
parallel algorithms for a large class of interesting problems.
Their original paper was
followed by a number of papers that consider the problem in other
shared-memory
models (see \cite{AndersonWoll,KS2,KPRS,KPS,MSP}).
 
The Write-All problem is, of course, a special case of the type of work
we consider.  Nevertheless, our framework differs from that of~\cite{KS}
in two important respects, so that their results do not apply to our
problem (nor ours to theirs).
First, they  consider the shared-memory model
while we consider the message-passing model.
Using the shared-memory model simplifies things considerably for our
problem.  In this model,
there is a straightforward algorithm
(that uses shared
memory to record what work has been done) with
optimal effort $O(n+t)$ (where effort now counts both reading and writing
into shared memory, as well as doing work),
running in time
$O(nt)$.\notefromO{replaced: $O(nt+t^2)$} While
there are well-known emulators that can translate algorithms from the
shared-memory model to the message-passing model (see
\cite{ABD,BD}), these emulators are not applicable for our problem,
because the number of failures they tolerate is less than a majority of the
total number of processes, while our problem allows up to $t-1$ failures. Also,
these transformations introduce a multiplicative overhead of message complexity
that is polynomial in $t$, while one of our goals here is to minimize this
term.\footnote{In fact, these emulators are designed for
asynchronous systems, and it may be possible to improve their
resilience for our synchronous model. Nevertheless,
they seem to have an inherent multiplicative overhead in message
complexity that is at least linear in $t$.}
Second, our complexity measure is inherently different from that of
\cite{KS}. Kanellakis and Shvartsman's complexity
measure is the sum, over the rounds during which the algorithm is running, of
the number of processes that are not faulty during each round.
They call their measure the {\em available processor
steps}.\notefromO{Added this sentence in order to use later when
talking about Yung et al.} This measure essentially ``charges''
for a nonfaulty process at round~$r$ whether or not it is actually doing
any work (say, reading or writing a cell in shared memory).
Our approach is generally {\em not} to charge a process in round
$r$ if it is not expending any effort (sending a message or
performing a unit of work) at that round, since it is free at that
round to be working on some other task.%
\footnote{Inactive processes in our algorithms may need to
both receive messages and
count the number of rounds that have
passed, say from the time they received
their last message.
We assume that processes can do this while carrying on other tasks.}
Of course, the appropriateness of charging or not charging for
process idle time will depend very much on the details of the
system and the tasks being performed.

Our results have been extended by De Prisco, Mayer, and
Yung~\cite{DMY94}, and by Galil, Mayer, and Yung~\cite{GMY95}.
De Prisco, Mayer, and Yung~\cite{DMY94} consider the problem
introduced here, but their goal is to optimize the available
processor steps defined by \cite{KS}, and then the number of
messages. They present a message-efficient algorithm that achieves
optimality in the available processor steps measure. They also
show that when $t \approx n$, any algorithm for performing work in
the message-passing model requires at least $n^2$  available
processor steps.
This lower bound can be avoided in shared-memory models that
allow concurrent writes; for example, an $O(n \log^2 n)$ solution
is presented in \cite{KS}.
Galil, Mayer, and Yung~\cite{GMY95}
employ the results of
\cite{DMY94} to obtain
a Byzantine agreement algorithm for
the crash fault model that requires only a linear number of
messages.
Roughly speaking, the processors are organized into a tree.
The children of the root attempt to solve the problem recursively;
the group membership protocol of~\cite{DMY94} (called a
{\em checkpoint protocol}, not to be confused with the checkpoints
as defined in our paper) is used to attempt to determine
which of the children have failed to complete the recursive step,
and the computation is re-organized accordingly.

The Galil, Mayer, and Yung protocol compares to that obtained
by using our Protocol~$\B$ as follows:
\cite{GMY95} requires $O(n)$ messages, while ours
requires $O(n \sqrt n)$;
\cite{GMY95} requires
$O(n^{1 + 1 / \epsilon})$ rounds of communication while ours
requires $O(n)$;
finally, \cite{GMY95} requires messages of length $\Omega(n + \log_2|V|)$,
where $V$ is the set of possible agreement values.
This appears to be because the protocol requires knowledge of
which processors are alive and which processors occupy which
parts of the tree.
In contrast, our messages are of length $O(\log n + \log_2|V|)$.

\notefromO{Maybe we should
omit this sentence or rewrite it somehow. Theirs uses
$n^{1+\epsilon}$ rounds and high bit complexity. In contrast to
ours, it does not simply use a work performing algorithm as a black
box (I think). They mention that their messages can be encoded in
order to get optimal bit complexity with the price of exponential
time or something. Details are in page 726 of FOCS '95.}

\section{A Protocol with Effort
\protect{$O(n+t^{3/2})$}}
 
Our goal in this section is to present a protocol with effort $O(n+t\sqrt{t})$
and running time
$O(n + t)$.  We begin with a protocol that
is somewhat simpler to present and analyze, with effort $O(n+t\sqrt{t})$
and running time $O(nt + t^2)$.  This protocol has the additional
property
of working with minimal change in an asynchronous environment with
failure detection.
 
The main idea of the protocol
is to use checkpointing in order to avoid
redoing too much work if a process fails.
The most na\"{\i}ve approach to checkpointing does not work.  To
understand why, suppose a process does a checkpoint after
each $n/k$ units of work. This means that up to
$n/k$ units of work are lost when a process fails.  Since
up to $t$ processes may fail, this means that $nt/k$ units
of work can be lost (and thus must be repeated), which
suggests we should take $k \ge t$ if we want to do no more
than $O(n)$ units of work altogether.  However, since each
checkpoint involves $t$ messages, this means that roughly
$tk$ messages will be sent.  Thus, we must have $k \le
\sqrt{t}$ if we are to use fewer than $t\sqrt{t}$
messages. Roughly speaking, this argument shows that doing
checkpoints too infrequently means that there might be a
great deal of wasted work, while doing them too often
means that there will be a great deal of message overhead.
Our protocol avoids these problems by doing {\em full}
checkpoints to all the processes relatively
infrequently---after $n/\sqrt{t}$ units of work---but
doing partial checkpoints to only $\sqrt{t}$ processes
after every $n/t$ units of work.  This turns out to be
just the right compromise.

\subsection{Description of the Algorithm}
For ease of exposition, we
assume that $t$ is a perfect square, and that $n$ is divisible by
$t$
(so that, in particular, $n > t$).
We leave to the reader the easy modifications of the protocol when
these assumptions do not hold.
We assume that the processes are numbered 0
through $t-1$, and that the units of work are numbered 1
through $n$.  We divide the processes into $\sqrt{t}$
groups of size $\sqrt{t}$ each, and use the notation $g_i$
to denote process $i$'s group. (Note $g_i = \lceil (i + 1)
/ \sqrt{t} \rceil$.) We divide the work into
$\sqrt{t}$ {\em chunks\/}, each of size $n/\sqrt{t}$, and subdivide the
chunks into
$\sqrt{t}$ {\em subchunks\/} of size $n/t$.
 
The protocol
guarantees that at each round, at most one process is {\em
active}.  The active process is the only process
performing work.  If process $i$ is active, then it knows
that processes 0 to $i-1$ have crashed or terminated.
Initially, process 0 is active. The algorithm for process
0 is straightforward: Process 0 starts out doing the work,
a subchunk at a time.  After completing a subchunk $c$, it
does a checkpoint to the remaining processes in its group
$g_0$ (processes $1$ to $\sqrt{t}-1$); that is,
it informs
its group that the subchunk
of work has been completed by
broadcasting to the processes in its
group a
message of the form
$(c)$.
(If process 0 crashes in the middle of a broadcast, we assume only
that some subset of the processes receive the message.)
We call this
a {\em partial checkpoint},
since the checkpointing is only to the processes in
$g_0$.
(Code for this for module and the whole protocol may be found in
Figure~\ref{protocol A code}.)
After
completing a whole
chunk of work---that is, after completing a subchunk $c$ which is a
multiple of $\sqrt{t}$---process 0 informs
{\em all\/} the processes that subchunk $c$ has been completed,
but it informs them one group at a time.  After informing
a whole group, it checkpoints the fact that a group has been informed to
its own group (i.e., group 1).  Formally, after
completing a subchunk $c$ that is
a multiple of $\sqrt{t}$, process 0 does a partial checkpoint to its own
group, and then for each group $2, \ldots, \sqrt{t}$, process 0
broadcasts to the
processes in group $g$ a message of the form $(c,g)$, and then
broadcasts to
all
the processes in its own group a message of the form $(c,g)$.  We call
this a {\em full checkpoint}.
Note
that in a full checkpoint,
there is really a double checkpointing process:  we
checkpoint both the fact that work has been completed, and (to the
processes in $g_0$) the
fact that all processes have been informed that the work has been
completed.
Process 0 terminates after sending the message $(t,\sqrt{t})$ to process
$t-1$,
indicating to the last process that the last chunk of work has been
completed
(unless it crashes before that round).
 
If process 0 crashes, we want process 1 to become active; if process 1
crashes, we want process 2 to become active, and so on.  More generally,
if process $j$ discovers that the first $j-1$
processes have crashed, then it becomes active.  Once process $j$ becomes
active, it continues with essentially the same algorithm as process 0,
except that it does not repeat the work it knows has already been done.
We must ensure
that the takeover proceeds in a ``smooth'' manner, so that there is at
most one active process at a time.

\begin{figure*}
\begin{tabbing}
XX \= XXXX \= XXXX \= XXXX \= XXXX \= XXXX \= XXXX \kill

Main protocol\\ \\
1. \> {\bf if} round number $=DD(j)$ {\em and} not received
   $(t)$ or $(t,g_j)$ \\
2. \> {\bf then} \>   DoWork.
\\ \\
DoWork\\
\\
1. \> {\bf if}\notefromO{Replaced the previous figure of protocol
\A~with two figures, in order to be able to use the new
Figure~\ref{dowork code} in the code for Protocol \B.}
the last message  received was from
$k$ and had
       the form $(c,g)$\\
2. \>{\bf then} \> {\bf if} $k \notin g_j$\\
3. \>\>{\bf then} \> Partialcheckpoint($c$);\ \ \  \{see code below\}\\
4. \>\> {\bf else} \> Broadcast $(c,g)$ to processes
$j+1,\ldots,g_j\sqrt{t}-1$;\\
5. \>\>\> Fullcheckpoint$(c,g+1)$;\quad \{complete a full checkpoint;
   see code below\}\\
6. \>{\bf else} \>let ($c$) be the last message received;\\
7. \>\> Partialcheckpoint($c$);\\
8. \>\>{\bf if} $c$ is a multiple of $\sqrt{t}$ \\
 
9. \>\>{\bf then} \>Fullcheckpoint($c, g_j+1$).\\
10. {\bf for} $s=c+1$ to $t$ {\bf do} \quad \{proceed with performing
                                              work\}\\
11. \>\>Perform subchunk $s$;\\
 
12. \>\>Partialcheckpoint($s$);\\
 
13. \>\>{\bf if} $s$ is a multiple of $\sqrt{t}$; \\
 
14. \>\>{\bf then} \>Fullcheckpoint($s, g_j+1$)\\ \\
Partialcheckpoint$(c)$\\
\\
1. \>Inform the remainder of group $g_j$ that subchunk $c$ has been
performed\\
\>by broadcasting $(c)$ to processes $j+1, \ldots,
g_j\sqrt{t}-1$\\
 
\\
\\
\>Fullcheckpoint$(c,l)$\\
\\
1. \>{\bf for} $g = l$ to $\sqrt{t}$ {\bf do}\\
 
2. \>\>Inform group $g$ that subchunk $c$ has been performed\\
   \>\>by broadcasting $(c,g)$ to group $g$;\\
 
3. \>\>Inform the remainder of group $g_j$
that group $g$ has been informed\\
\>\> about subchunk $c$ by broadcasting
$(c,g)$ to processes $j+1,\ldots,g_j\sqrt{t}-1$
\end{tabbing}
\caption{Protool $\A$; Code for Process $j$}
\label{protocol A code}
\end{figure*}
 
Process $j$'s algorithm
is as follows. If $j$ does not know that all
the work has already been performed and sufficiently long time has passed from
the beginning of the execution, then $j$ becomes
active.  ``Sufficiently long'' means long enough to ensure that processes
$0,\ldots,j-1$ have crashed or terminated. As we show below, we can take
``sufficiently long'' to be defined by the function
$DD(j)=j(n+3 t)$.
(``DD'' stands for deadline.
We remark that this is not an optimal choice for the deadline;
we return to this issue later.)
Thus,
if the round number $r$ is less than
$DD(j)$, then $j$ does nothing.  Otherwise, if $j$ does not
know that the work is completed,
it takes over as the active process at round
$DD(j)$.

\remove{orli:

\begin{figure*}
\hrule
\begin{tabbing}
XXXXX \= XXXX \= XXXX \= XXXX \= XXXX \= XXXX \= XXXX
\kill
 
1. \>If round number $=DD(j)$ {\em and} \\
   \>you have not received
   $(t)$ or $(t,g_j)$ then \\
 
2. \>\>If the last message you received was from $k$ and had
       the form $(c,g)$ then\\
3. \>\>\> If $k \notin g_j$ then \\
4. \>\>\>\> Partialcheckpoint($c$)\\
5. \>\>\> If $k \in g_j$ then \\
6. \>\>\>\> Broadcast $(c,g)$ to processes $j+1,\ldots,g_j\sqrt{t}-1$\\
\\
   \>\>\>\> \{Complete a full checkpoint\}\\
7. \>\>\>\> Fullcheckpoint$(c,g+1)$\\
\\
 
8. \>\>Else let ($c$) be the last message received;\\
   \>\>\> Partialcheckpoint($c$)\\
9. \>\>\>If $c$ is a multiple of $\sqrt{t}$ then \\
 
10. \>\>\>\>Fullcheckpoint($c, g_j+1$).\\
\\
 
\>\>\{Proceed with performing work\}\\
 
11. \>\>For $s=c+1$ to $t$ do\\
 
12. \>\>\>Perform subchunk $s$\\
 
13. \>\>\>Partialcheckpoint($s$)\\
 
14. \>\>\>If $s$ is a multiple of $\sqrt{t}$ then \\
 
15. \>\>\>\>Fullcheckpoint($s, g_j+1$).\\
\end{tabbing}
\caption{Protool $\A$; Code for Process $j$}
\hrule
\label{protocol A code}
\end{figure*}

}

When $j$ takes over as the active process, it essentially follows
process 0's algorithm.  Suppose the last message $j$ received was
of the form $(c,g)$, and this message was received from a process $k$.
By the syntax of the message we have that $c$ is a multiple of $\sqrt{t}$
and that $k$ was performing a full checkpoint when it sent the message.
If $k \notin g_j$ then $g = g_j$, since this is the only kind of
full checkpoint message that $k$ sends to processes outside its
group.  Thus, $j$ must inform the rest of its own group that
subchunk $c$ was performed, which it does with a Partialcheckpoint($c$),
and proceeds with the full checkpoint of $c$, beginning with group
$g_i+1 = g+1$.
 
If $k \in g_j$ then $g > g_j$; the meaning of $(c, g)$ in this case
is that $k$ has told group $g$ that subchunk $c$ has been completed,
and is telling its own group, $g_k$ ($ = g_j$), about this fact.
In this case $j$ first ensures that its own group knows that
group $g$ has been informed about subchunk $c$ by broadcasting $(c,g)$
to the remainder of its group, and then proceeds with the full checkpoint
beginning with group $g+1$.
 
If the last message received was of the form $(c)$ then
this message was part of a partial checkpoint to $g_j$.  In this
case process $j$ completes the partial checkpoint.
 
In all three cases, $j$ proceeds with work beginning with
subchunk $c+1$ (if such a subchunk exists).
 
Unless it has already crashed,
process $j$ terminates before becoming active if it receives
$(t)$ (as part of a partial checkpoint)\notefromO{removed; I don't
understand why this is true. `; this only occurs if $g_j =
\sqrt{t}$} or
$(t,g_j)$ (as part of a full checkpoint).\notefromO{removed: `;
this may occur whether or not $g_j =
\sqrt{t}$'} Otherwise, after becoming active at $DD(j)$, it
terminates as follows.  If $g_j = \sqrt{t}$ then $j$ terminates
after broadcasting $(t)$ to the remainder of $g_j$.
\remove{orli11: If $g_j < \sqrt{t}$ then $j$ terminates after
completing a call of the form Fullcheckpoint($t, l$) for some $l$.}
If $g_j < \sqrt{t}$ then $j$ terminates after
completing a call of the form Fullcheckpoint($t, g_j+1$).
This completes
the description of our first protocol.
We call this Protocol~$\A$;
the code appears in Figure~\ref{protocol A code}.

Notice that we can easily modify this algorithm to run in a completely
asynchronous system
equipped with an appropriate {\em failure detection mechanism}
\cite{CT}:  If a process fails,
then the failure detection mechanism must
eventually inform all the processes that have not failed of this
fact; moreover, the mechanism must be sound, in that it never
says that a nonfaulty process has failed.
\notefromO{Added end of sentence. Not
sure what else the referee wants. Maybe he wants references to
failure detection mechanisms?}  The modification is trivial:
rather than waiting until round $DD(j)$ before becoming active,
process $j$ waits until it has been informed that processes $1,
\ldots, j-1$ crashed or terminated.
 
\subsection{Analysis and Proof of Correctness}
\remove{orli11: the paragraph below sounds starnge. It made sense in the
abstract, but in the full paper, of course we give a complete proof of
correctness.}
We now give a correctness proof for Protocol ${\cal A}$.
We say a
process is {\em retired\/} if it has either crashed or terminated.
 
\begin{lemma}\label{facts}
A process
performs at most $n$ units of work, sends at most $3t \sqrt{t}$
messages, and runs for less than $n+3t$ rounds from the time it
becomes active to the time it retires.
\end{lemma}
 
\begin{proof}
It is easy to see that from the time process $i$ becomes active,
it performs each unit of work
at most once, partial checkpoints each subchunk at most once (and
hence performs at most $t$ partial checkpoints), and full checkpoints
every chunk at most once (and hence performs at most $\sqrt{t}$ full
checkpoints).  Each partial checkpoint consists of
a broadcast to process $i$'s
group, and hence involves at most $\sqrt{t}$ messages and one round.
Thus, process $i$ spends at most $t$ rounds on partial checkpoints,
and sends at most $t \sqrt{t}$ messages when performing partial
checkpoints.  During a full checkpoint, process $i$
broadcasts once to each group other than
its own, and broadcasts at most $\sqrt{t}$ times to its own group.
Each broadcast involves at most $\sqrt{t}$ messages and one round, and
there are $\sqrt{t}$ groups.  Thus, process $i$
sends less than
$2t \sqrt{t}$ messages
when performing full checkpoints, and takes less than
$2t$ rounds doing so.  The required bounds immediately follow.
\end{proof}

Recall that $DD(j) = j(n+3t)$.
The following lemma is now immediate from the definition of DD.
\begin{lemma}
\label{ordergroupALemma}
Assume process $j$ becomes active at round $r$
of an execution $e_{\cal A}$ of
Protocol $\A$. Then all processes $<j$ have
retired before round~$r$. \QED
\end{lemma}
 
\remove{
\begin{proof}
Since $i$ follows Protocol $\A$, $r$ must be equal to $i
(\maxactivity)$. However, each process $j<i$ that ever becomes active,
does so by time $j (\maxactivity)$, and hence by
Fact 3 must have retired by time $i (
\maxactivity)$.
\end{proof}
}

It is sometimes convenient to view a group $g_i$
as a whole. Therefore we say that a {\em group is
active} in the period starting when some process in this group becomes
active and ending when the last process of this group retires.
Notice that
Lemma~\ref{ordergroupALemma} ensures that when $g_i$ becomes active, all
processes in smaller groups have retired.
 
\begin{theorem}
\label{protocolATheorem}
In every execution of Protocol $\A$,
\begin{itemize}
\item[(a)]
at most
$3n$ units of work are performed in total by the processes,
\item[(b)]
at most $9 t \sqrt{t}$ messages are sent,
\item[(c)] by round $nt+3t^2$, all processes have retired.
\end{itemize}
\end{theorem}
 
\begin{proof}
Part (c) is immediate from
\notefromO{replaced: Lemma~\ref{ordergroupALemma}}
Lemma~\ref{facts} and the definition of
$DD$.
 
We prove parts (a) and (b) simultaneously.
To do so, we need a careful way of counting the total number of
messages sent and the total amount of work done.  A given unit
of work may be performed a number of times.  If it is performed
more than once, say by processes $i_1, \ldots, i_k$, we say that
$i_2$ {\em redoes\/} that unit of work of $i_1$, $i_3$ redoes the work of
$i_2$, etc.  It is important to note that $i_3$ does not redo the work
of $i_1$ in this case; only that of $i_2$.  Similarly, we
can talk about a message sent during a partial checkpoint of
a subchunk
or a full checkpoint of a chunk done by $i_1$ as being
{\em resent\/} by $i_2$.
In particular, a message $m$ sent by $i_1$ as part of a
broadcast is resent by $i_2$ if $i_2$ sends exactly the same
message as part of a broadcast (not necessarily to the same
set of recipients).
For example, if $i_1$ sends $(c)$ to the remainder of
$g_{i_1}$ as part of
a partial checkpoint, and later $i_2$ sends $(c)$ to
the remainder of $g_{i_2}$,
then, whether or not $g_{i_1} = g_{i_2}$, the messages in the
second broadcast are considered to be resendings.
\remove{Lemma~\ref{wastegroupsLemma}}
 
Since the completion of a chunk is followed by a full checkpoint,
it is not hard to show that when a new
group becomes active,
it will redo at most one chunk of work that was
already done by previous active groups.  It will also
redo at most one full checkpoint
that was done already on the previous chunk, and $\sqrt{t}$
partial checkpoints (one for each subchunk of work redone).
\notefromO{removed since I think it is included already. `Finally,
if
$g_j < g_i$, and the last message sent by process $j$ before
crashing is a broadcast to process $i$'s group that was not
received by $i$, process $i$ must
resend this broadcast.'}  In all, it is easy to see that at most
$n/\sqrt{t}$ units of work done by previous groups
are redone when a new group becomes active,
and $3t$ messages are resent.
Similarly,
since the completion of a subchunk is followed by a partial checkpoint,
it is not hard to show
\remove{In addition,
Lemma~\ref{wastegroupLemma} implies}
that when a new
process, say $i$,
in a group that is already active becomes active, and the last
message it received was of the form
$(c)$ (i.e., a partial checkpoint of subchunk $c$), it will
redo at most
one subchunk that
was already done by previous active process (namely, $c+1$), and may possibly
resend the messages in two partial checkpoints:~the one sent
after subchunk $c$, and the one sent after subchunk $c+1$ (if the
previous process crashed during the checkpointing of $c+1$ without $i$
receiving the message).
If the last message that $i$ received was $(c,g)$ for $g > g_i$
(that is, the checkpointing of a checkpoint in the middle of
a full checkpoint), then similar arguments show that it may resend
$3\sqrt{t}$ messages:~the checkpoint of $(c,g)$ to its own group,
the checkpoint $(c,g+1)$ to group $g+1$, and the checkpointing of
$(c,g+1)$ to its own group.
Thus, the amount of work done by an active group that is
redone when a new process in that group becomes active
is
at most $n/t$, and the number of messages resent is at most $3\sqrt{t}$.
 
The maximum amount of unnecessary work done is:
(number of groups) $\times$ (amou\-nt of work redone when a new group
becomes active) + (number of processes) $\times$ (amount of work
redone when a new process in an already active group becomes active)
$\le  \sqrt{t}(n/\sqrt{t}) + t(n/t) = 2n$.
Similarly,
the maximum number of unnecessary
messages that may be sent is
no more than: (number of groups) $\times$ (number of messages
resent when a new
group becomes active) + (number of processes) $\times$ (number
of messages resent when a new process in an already active
group becomes active) $\le
\sqrt{t}(3t)+
t 3\sqrt{t}=6t \sqrt{t}$.
Clearly $n$ units of work must be done; by Lemma~\ref{facts},
at most $3t\sqrt{t}$ messages are necessary.
Thus, no more than $3n$ units of work will be done altogether, and
no more than
$9t
\sqrt{t}$ messages will be sent altogether.
\end{proof}

\subsection{Improving the Time Complexity}
 
As we have observed, the round complexity of Protocol $\A$~is $nt+3t^2$.  We now
discuss how the protocol can be modified to give a
protocol that has round complexity $O(n + t)$,
while not significantly changing the amount of work done
or the number of messages sent.
 
Certainly one obvious hope for improvement is to use a better function
than $DD$ for computing when process $i$ should become active.  While some
improvement is possible by doing this, we can get a round complexity of
no better than $O(n\sqrt{t})$ if this is all we do, which is still more than we
want. Intuitively, the problem is that if process $j$ gets a message of the form
$(c,g)$, then it is possible, as far as $j$ is concerned, that some other process
$i<j$ may have received a message of the form $(c+\sqrt{t},h)$. (Observe that
this situation is possible even if $g_i=g_j$ because if the sender of the
message $(c+\sqrt{t}, h)$ crashes at the round it broadcasts this message to
$g_i$, this message may reach an arbitrary subset of the processes in $g_i$.)
Process $j$ cannot become active before it is sure that $i$ has retired. To
compute how long it must wait before becoming active, it thus needs to compute
how long $i$ would wait before becoming active, given that $i$ got a message of
the form $(c+\sqrt{t},h)$. On the other hand, if $i$ did get such a message,
then as far as $i$ is concerned, some process $i' < i$ may have received a
message of the form $(c+2\sqrt{t},h')$. Notice that, in this case, process $j$
knows perfectly well that no process received a message of the form
$(c+2\sqrt{t},h')$; the problem is that $i$ does not know this, and
must take into account this possibility when it computes how long to wait
before becoming active. Carrying out a computation based on these
arguments gives an algorithm which runs in $O(n\sqrt{t})$ rounds.
 
On closer inspection, it turns out that the situation described
above really causes difficulties only when all processes involved (in the
example above, this would be the processes $j$, $i$, and $i'$) are in
the same group. Thus, in our modified algorithm, called Protocol $\B$, process
$j$ computes the time to become active as follows: Suppose that the
last message received by process $j$ before round $r$ was received from
process $i$ in round $r'$. Process $j$ then computes a function $\DDB(j,i)$
with the property that if $r = r' + \DDB(j, i)$, then process
$j$ knows at round $r$ that all processes in groups $g' < g_j$ must have
retired. Moreover, if $g_i=g_j$, then $j$ knows at round $r$ that all
processes $\le i$ must have retired. Process $j$ then polls all the
lower-numbered processes in its own group not known to it as retired, one
by one, to see if they are alive; if not, then $j$ becomes active. If any of
them is alive, then the lowest-numbered one that is alive becomes active upon
receipt of $j$'s message.
Once a process becomes active, it proceeds just as in
Protocol~$\A$. This technique turns out to save a great deal of time, while
costing relatively little in the way of messages.
 
In particular, in Protocol $\B$, process 0 follows the same algorithm as in
Protocol $\A$. Process $j$'s algorithm is as follows.
Here $j$ receives messages
either of the form $(c)$,
$(c,g)$ or of the form \goahead. We call the first
two types
of messages {\em ordinary}, to distinguish them from the
\goahead~messages. Suppose that the last ordinary message received
by process $j$ before round $r$ is
of the form $(c)$ or
$(c, g)$, and this message was received from process
$i$ at round $r'$.
To avoid
dealing separately with the special case in which $j$ does not receive any
message before it becomes active, we use the convention that process 0
becomes active in round 0 (just before the execution begins) and every process
receives from it an ordinary message $(0, g)$ at that round. (These fictitious
messages  are used only in the analysis and hence will not be taken into account
when computing the message complexity of the protocol. Also, if in the actual
execution process 0 crashes before ever becoming active then we say that it
crashes just after it finishes broadcasting these fictitious messages.)
There are now two ways for $j$ to become active at round $r$. The first is if $j$
receives a \goahead~message at round $r$ and $c<t$. In this case $j$ becomes
active, proceeding just as in Protocol \A~when it became active at
round $DD(j)$. Alternatively, if $j$ does not receive a message for a
sufficiently long time, $j$ becomes active. Intuitively, sufficiently long
will ensure that all processes smaller than $j$ have already retired.

To analyze this more formally, we need some definitions.
Let \mpd~be $n/t+2$.
(``\mpd'' stands for process time out.)
$\mpd-1$
is an upper bound on the
number of rounds that can pass before process $j$ in group $g_j = g_i$
hears from process $i$ if $i$ is active.  Let $\overj$
denote $j$ mod $\sqrt{t}$.\notefromO{defined separately in
response to referee.} Let $\mgd(i)$ be
$n/\sqrt{t}+3\sqrt{t}+(\sqrt{t}-\overi-1)\mpd+1$. (``\mgd'' stands
for group time out.)
$\mgd(i)-1$ is
an upper bound on the
number of rounds that can pass before
process $j$ in group $g_j > g_i$ hears from a process $k$ in $g_i$
with $\overi \le \overk
\le \sqrt{t}-1$ if any of these processes is active. Next we define a
new deadline function as follows:
 
$$\DDB(j,i)=\left\{
\begin{array}{ll}
\mgd(i)+(g_j-g_i-1)\mgd(0)&
\mbox{if $g_j \ne g_i$} \\
\mpd &  \mbox{otherwise.}
\end{array}
\right.$$
We now define ``sufficiently long'' in terms of $\DDB$ rather than $DD$.
Again taking $r$ to be the current round, $r'$ to be the last round
before $r$ in which $j$ received a message, and $i$ to be the process
sending that message, $j$ proceeds as follows:
If $r < r'+\DDB(j,i)$, then $j$ does nothing.  If $c < t$ and $r =
r'+DD_B(j,i)$, $j$ becomes {\em preactive}. First consider the case where $g_i
\ne g_j$. Informally, at this point, $j$ knows that all processes from groups $<
g_j$ have failed. In this case, it sends a \goahead~message to each
lower-numbered process in its group, starting with the first
process in $g_j$ up to the $(\overj-1)$st process in $g_j$, and waiting
$\mpd-1$ rounds between messages to see if it receives a message. (Observe that
if the recipient of the \goahead~message is alive, the sender receives a
message from it within one round after the \goahead~mesage was sent; however, for
technical reasons the sender of the \goahead~messages waits $\mpd-1$
rounds between two successive \goahead~messages.)
If $g_i = g_j$, process $j$ proceeds similarly to the
case where $g_i \ne g_j$ except that when sending \goahead~messages it starts
with the $(\overi+1)$st process in $g_j$. That is, it sends a \goahead~message to
each lower-numbered process in its group, starting with the $(\overi+1)$st
process in $g_j$ up to the $(\overj-1)$st process in $g_j$, and waiting $\mpd-1$
rounds between messages to see if it receives a message. If $j$ does not receive
any response to its \goahead~messages by round $r'+\DDB(j, i) + \overj \mpd-1$
if $g_i \ne g_j$, or by round $r' + (\overj-\overi) \mpd-1$ if $g_i=g_j$, then
it becomes active at round $r' + \DDB(j, i)+\overj \mpd$ (respectively, $r' +
(\overj-\overi)\mpd$), proceeding just as it did in Protocol $\A$ at round
$\DD(j)$. If it does get a message, then $j$ becomes passive again.

\begin{figure*}[htb]
\begin{tabbing}
XX \= XXXX \= XXXX \= XXXX \= XXXX \= XXXX \= XXXX
\kill

Main protocol\\ \\
1.  \>{\bf if}\notefromO{Added this figure.} just received a
\goahead~message\\
2. \>{\bf then} \>DoWork; \ \ \ \ \{see Figure~\ref{protocol A code} for
      details\}\\
3. \>{\bf else} \> {\bf if} round number $= r'+\DDB(j, i)$\\
\> \>  {\em and\/} last message received was from process $i$ at round $r'$\\
 
4. \>\>{\bf then} \>PreactivePhase$(i, r')$\\ \\
\\
 
PreactivePhase$(i, r')$\\ \\
1. \>{\bf if} \notefromO{Added the code for protocol \B}
$g_i \ne g_j$ \\

2.  \>{\bf then} \>$i' \gets (g_j-1) \sqrt{t}$;\\
3. \>{\bf else} \>$i' \gets i+1$;\\
 
\\
4.  \>$r \gets \curr$;\\
5.  \>{\bf while} not received a message {\em and\/} $i' < j$
{\bf do}\\
6.  \>\> {\bf if} $\curr - r \equiv 0 \ (\mod PTO)$\\
 
7. \>\>{\bf then} \> send \goahead~message to process $i'$;\\
8. \>\> \> $i' \gets i'+1$;\\

9.  \>{\bf if} just received a \goahead~message or $i'\ge j$\\
 
10. \>{\bf then} \>DoWork
 
\end{tabbing}
\caption{Protool $\B$; Code for Process $j$}
\label{protocol B code}
\end{figure*}

Note that the construction of the algorithm guarantees that if $r
= r'+\DDB(j, i)+\overj \mpd$, the last ordinary message that $j$
receives in execution
$e_{\cal B}$ prior to round $r$ was sent by
$i$, and $g_i \ne g_j$,
then $j$ will
become active in $e_{\cal B}$ at or before round $r$. (It may become active
earlier if it receives a \goahead~message.) Similarly, if
$r = r'+(\overj-\overi) \mpd$,\notefromO{replaced: $r = r'+\DDB(j,
i)+(\overj-\overi) \mpd$}
the last ordinary message that $j$
receives in execution $e_{\cal B}$ prior to round $r$ was sent by
$i$, and $g_i = g_j$, then
$j$ will become active in $e_{\cal B}$ at or before round $r$. Define
$$\Tr(j,i)=\left\{
\begin{array}{ll}
\mgd(i)+ (g_j-g_i-1)\mgd(0)+\overj \mpd&
\mbox{if $g_j \ne g_i$} \\
(\overj - \overi)\mpd &  \mbox{otherwise.}
\end{array}
\right.$$
(``\Tr''stands for transition time.) Our observations above show that if $r =
r'+\Tr(j, i)$ and the last ordinary message that $j$ receives in execution
$e_{\cal B}$ prior to round $r$ was sent by $i$, then $j$ will become active in
$e_{\cal B}$ at or before round $r$.
 
Unless it has already crashed,
process $j$ terminates before becoming active if it receives
$(t)$ (as part of a partial checkpoint)\notefromO{removed again:
`; this only occurs if $g_j = \sqrt{t}$'} or $(t,g_j)$ (as part of
a full checkpoint).\notefromO{Again: `; this may occur whether or
not $g_j = \sqrt{t}$'} Otherwise, after becoming active it
terminates as follows.  If $g_j = \sqrt{t}$ then $j$ terminates
after broadcasting $(t)$ to the remainder of $g_j$.
\remove{orli11: If $g_j < \sqrt{t}$ then $j$ terminates after
completing a call of the form Fullcheckpoint($t, l$) for some $l$.}
If $g_j < \sqrt{t}$ then $j$ terminates after
completing a call of the form Fullcheckpoint($t, g_j+1$). This
completes the description of Protocol $\B$. The\notefromO{Added
rest of subsection.} code for Protocol~$\B$
appears in Figure~\ref{protocol B
code};
it uses the code for the DoWork procedure in Figure~\ref{protocol A
code}.

\subsection{Proof of Correctness of Protocol $\B$}
\label{section: Bproof}
 
In this section we show that the round complexity of Protocol $\B$~is
$O(n+t)$, and that neither the amount of work done nor the number of messages
sent in Protocol \B~is significantly larger than in Protocol \A.
 
Suppose for a moment that in every execution of Protocol $\B$~a
process becomes active only after all lower numbered processes have retired.
Since when a process becomes active in an execution of Protocol $\B$ it
performs essentially the same steps as when it becomes active when it follows
Protocol \A, a similar proof to the one of Theorem~\ref{protocolATheorem} will
show that the amount of work performed in any execution of Protocol $\B$ is no
more than $3n$ units (which is roughly the maximum amount of work performed in
any execution of Protocol \A), and the number of ordinary messages sent is no
more than $9t\sqrt{t}$ (which is roughly the maximum number of messages sent in
any execution of Protocol \A). Since the number of \goahead~ messages sent in
any execution of Protocol \B~is at most $t \sqrt{t}$ (each process sends at most
one \goahead~message to every other process in its group), it follows immediately
that the total amount of effort performed by Protocol \B~is not significantly
larger than the one performed by Protocol \A. Therefore, the main property we
need to prove is that in every execution of Protocol \B, a process becomes active
only after all lower numbered processes have retired.
 
Our analysis uses what we call {\em activation chains}. The
round $r$
{\em activation
chain\/} of process $i$,
denoted $\ac(i,r)$,
is the sequence of processes $\langle i_m, \ldots, i_0 \rangle$ such
that $i_0 = i$ and for all $j$,
if $i_j$ received an ordinary message prior to round
$r$, then $i_{j+1}$ is the sender of the last
message received by $i_{j}$.  (As we show below, it cannot be the case
that $i_j$ receives ordinary messages from two distinct processes
in the same round.  Since we have not yet proved this, for now, if
$i_j$ received ordinary messages from more than one process in the
last
round
in which
it received an ordinary message, we take $i_{j+1}$ to be the
lowest-numbered process among them.)
Notice that our convention that process 0 sent a round 0 message guarantees
that $i_m = 0$.
In
addition, note that the processes in the activation
chain appear in increasing
order since a process sends messages only to higher numbered processes.

It is sometimes
convenient to view $i$'s activation chain as
a whole and to reason about the effort performed by the chain. We say
that process $k$ in
$\ac(i,r)$
is the {\em current process} from
the round it becomes active up to (but not including) the round at which its
successor in the chain becomes active. Note that a process
in $\ac(i,r)$ is
current when it first becomes active. Now let $k, l$ be processes
in $\ac(i,r)$
such that $l$ immediately succeeds $k$, and assume the last
ordinary message $l$ receives from $k$ before $l$ becomes active is sent at
round $r'$. Clearly, any operation (sending messages or performing a unit of
work) done by $k$ after round $r'$ is not known by $l$ and hence may be repeated
by the chain (that is, may be repeated by some process when it is the current
process in the chain). On the other hand, any operation done by $k$ before round
$r'$ will be known by the processes succeeding it in the chain by the time
they become active, and hence will not be repeated by the chain. The
operation done by $k$ in round $r'$, which is a broadcast to $g_l$, will be
repeated by $l$ in the first round in which $l$ becomes active. We
say that an operation performed by a process $k$ in the chain is {\em useful} if
it is performed before the round in which the process immediately succeeding $k$
in the chain heard from $k$ for the last time before becoming active (if the
process is $i$, then there is no later process in the chain, and hence all
operations performed by $i$ are useful).
When we refer to
an
{\em operation performed
by a chain $\ac(l,r)$\/}, we mean
a useful operation performed by some process in that chain.
We say that a round is {\em useful\/} for the chain $\ac(l,r)$ if
the chain performed a useful operation in that round; otherwise
we say that the round is {\em useless}.
 
The discussion above shows that
the operations performed by a chain
proceed in a similar order to the operations performed by a single
active process. More precisely, if we consider only useful operations, the
processes in the
chain perform work units one by one in the natural order and without
repetition; each time a subchunk $c$ is completed by the chain, the group of the
process that completes this subchunk is informed
about this fact exactly once, and if the completed subchunk is a multiple of
$\sqrt{t}$, then in addition all groups whose numbers are higher than the group
of the process that completed this subchunk are informed that the subchunk is
completed one by one in the natural order and without repetition (that is, each
such group $g$ receives a message $(c, g)$ exactly once); moreover, each time
such a group $g$ is informed, this fact is checkpointed to the group of
the informer exactly once.
Assume process $i$ is active at some round $r$ with $r \ge r2  \ge r1
\ge 1$. Then we denote by $T^{r2}_{r1}(i)$ the number of
useful rounds for the chain $\ac(i,r)$ in
interval $[r1, r2]$ (that is, in the period from round $r1$ to round
$r2$ (inclusive)). The discussion above shows the following:
 
\begin{lemma}
\label{maxLemma}
Let $l$ be active at some round $r$ with $r \ge r2 \ge r1 \ge 1$. Then
\begin{itemize}
\item[(a)]
$T^{r2}_{r1}(l) \le n+3t$,
\item[(b)]
if $T^{r2}_{r1}(l) \ge n/\sqrt{t}+3 \sqrt{t}$, then each process
$\ge l$ must have received a message from
(a process in) $\ac(l,r)$
at some round
$r'$ such that $r1 \le r' \le r2$.
\end{itemize}
\end{lemma}
 
\begin{proof}
Part (a) follows from the fact that in each useful round, the chain either
performs work, or checkpoints to some group $g$ the fact that a subchunk $c$ was
performed, or checkpoints the fact that group $g$ was informed that chunk $c$
was performed. The discussion above shows that no unit of work is repeated
and hence there are at most $n$ useful rounds in which the chain performs work.
Similarly, each subchunk is partially checkpointed at most once and hence there
are at most $t$ useful rounds in which the chain performs partial checkpoints of
subchunks. Also, the completion of a chunk
is checkpointed to each group at most once, yielding at most $t$ useful rounds
in which such subchunks are checkpointed. Finally, the fact that group $g$
was informed about chunk $c$ is checkpointed at most once, yielding at most $t$
additional useful rounds. Summing the above the claim follows.
 
Part (b) follows because, as reasoned above, the
useful operations done by the chain follow the same order as if they are done by
a single active process, and hence within $n/\sqrt{t}+3 \sqrt{t}$ rounds the
chain must complete a chunk and a full checkpoint.
\end{proof}
 
Now, as we mentioned above, at the core of our proof of correctness is the fact that
when a process becomes active, all lower numbered processes have already
retired.
To prove this, we first prove a
lower bound on the number of useful rounds
for a given activation chain
in a given period.
Using this bound, we can show
that if some process $i$ receives its last ordinary
message before becoming active at round $r1$,
$i$
becomes active at
round $r2$, and some process $l < i$ has not retired by $r2$, then
process $i$ would have received an ordinary message from some process in
$\ac(l,r2)$ between rounds $r1$ and $r2$, contradicting our choice of
$r1$.

We now proceed with the formal proofs.
We start with a technical lemma.
\begin{lemma}
\label{addLemma}
Let $l>j>k$. Then
\begin{itemize}
\item[(a)]
$\Tr(j, k)+\Tr(l, j) = \Tr(l, k)$,
\item[(b)]
if $g_j < g_l$, then $\Tr(j, k)+\DDB(l, j) = \DDB(l, k)$.
\end{itemize}
\end{lemma}
 
\begin{proof}
The proof is straightforward. We start with Part (a).
In the calculations below, we use ``($g_i = g_j$)'' to denote
the value 1 if $g_i=g_j$ and 0 otherwise. Similarly, ``$(g_i \ne g_j)$'' denotes 1
if $g_i \ne g_j$, and 0 otherwise. Recall that $\overj$ denotes
$j$ mod $\sqrt{t}$.\notefromO{Added because of referee.}
\begin{eqnarray*}
\Tr(j, k)+\Tr(l, j)
&=&[\mdg(k)+(g_j-g_k-1)\mdg(0)+\overj \mdp](g_j \neq g_k)\\
&\quad&
+[(\overj-\overk)\mdp](g_j=g_k) \\
& \quad &+
[\mdg(j)+(g_l-g_j-1)\mdg(0)+\overl \mdp](g_l \neq g_j)\\
&\quad &
+[(\overl-\overj)\mdp](g_l=g_j).
\end{eqnarray*}
 
If $g_j=g_k$, then
\begin{eqnarray*}
\Tr(j, k)+\Tr(l, j)
&=&
(\overj-\overk)\mdp\\
&\quad &
+
[\mdg(j)+(g_l-g_k-1)\mdg(0)+\overl \mdp](g_l \neq g_k)\\
&\quad& +
[(\overl-\overj)\mdp](g_l=g_k)\\
&=&
[\mdg(k)+(g_l-g_k-1)\mdg(0)+\overl \mdp](g_l \neq g_k)\\
&\quad & +
[(\overl-\overk)\mdp](g_l=g_k)\\
&=&
\Tr(l, k),
\end{eqnarray*}
and part (a) follows.
(In the first equality we replaced $g_j$ by $g_k$ since in this case they are
identical, and the second equality follows because
$\mdg(j)+(\overj-\overk)\mpd=\mdg(k)$.)
 
If $g_j \ne g_k$, then
\begin{eqnarray*}
\Tr(j, k)+\Tr(l, j)
&=&
[\mdg(k)+(g_j-g_k-1)\mdg(0)+\overj \mdp]\\
&\quad & +
[\mdg(j)+(g_l-g_j-1)\mdg(0)+\overl \mdp](g_l \neq g_j)\\
& \quad &
+[(\overl-\overj)\mdp](g_l=g_j)\\
&=&
[\mdg(k)+(g_l-g_k-1)\mdg(0)+\overl \mdp](g_l \neq g_j)\\
&\quad & +
[\mdg(k)+(g_j-g_k-1)\mdg(0)+\overl \mdp](g_l = g_j)\\
&=&
[\mdg(k)+(g_l-g_k-1)\mdg(0)+\overl \mdp]\\
&=&
\Tr(l, k),
\end{eqnarray*}
and again part (a) follows.
(The second equality follows
by a case analysis on whether or not $g_l = g_j$, using the fact
that
$\mdg(j)+\overj \mpd=\mdg(0)$ and the
fourth equality follows since $g_j \ne g_k$ and $l>j>k$ implies $g_l \ne
g_k$.)
 
The proof of Part (b) is similar. Observe that here by assumption, $g_l \ne
g_j$ and hence also $g_l \ne g_k$. If $g_j=g_k$, then
\begin{eqnarray*}
\Tr(j, k)+\DDB(l, j)
&=&
(\overj-\overk)\mdp
+
[\mdg(j)+(g_l-g_k-1)\mdg(0)]\\
&=&
[\mdg(k)+(g_l-g_k-1)\mdg(0)]\\
&=&
\DDB(l, k),
\end{eqnarray*}
and part (b) follows.
 
If $g_j \ne g_k$, then
\begin{eqnarray*}
\Tr(j, k)+\DDB(l, j)
&=&
[\mdg(k)+(g_j-g_k-1)\mdg(0)+\overj \mdp]\\
&\quad & +
[\mdg(j)+(g_l-g_j-1)\mdg(0)]\\
&=&
[\mdg(k)+(g_l-g_k-1)\mdg(0)]\\
&=&
\DDB(l, k),
\end{eqnarray*}
and Part (b) follows.
\end{proof}
 
The next lemma establishes a lower bound on the number of useful rounds
for an activation chain in a given interval.
 
\begin{lemma}
\label{minLemma}
Assume $l$ is active at some round $r$ such
that $r \ge r2 \ge r1 \ge 1$. Assume $p \ge k$ is the current process in
$\ac(l,r)$
at some round $\le r1$. Then $T^{r2}_{r1}(l) \ge r2-r1-\Tr(l, k)+1.$
\end{lemma}
 
\begin{proof}
We first show that if $j$ is
in $\ac(l,r)$
and becomes active at round $r'$ with $r1 \le r' \le
r2$, then there are at most $\Tr(j, k)$ useless rounds in $[r1, r'-1]$. We
proceed by induction on $r'$. If $r'=r1$, the result is trivial. If $r' > r1$,
then $j$ is $i$'s successor for some $i$ in the activation chain
and $j$ received its last message from $i$ at some round $r''$. (There is such
an $i$ and such a message since by convention process 0 sent an ordinary message
to everybody just before the execution begins.)
By definition, we have $r' \le
r''+\Tr(j, i)$. If $i=k$, we are done, since no round in $[r1, r''-1]$ is
useless,
so there are at most $\Tr(j,k) = \Tr(j,i)$ useless rounds in
$[r1,r'-1]$.
If $i > k$, then suppose $i$ becomes active at $r'''$. By the
inductive hypothesis, there are at most $\Tr(i, k)$ useless rounds in $[r1,
r'''-1]$. All the rounds in $[r''', r''-1]$ are useful. Thus, there are at most
$\Tr(j, i)+\Tr(i, k)$ useless rounds in $[r1, r'-1]$. Since $\Tr(j, i)+\Tr(i,
k)=\Tr(j, k)$ by Lemma~\ref{addLemma}, the inductive step follows.

Suppose that $l$ becomes active at round $r3$.
By the argument above, there are at most $\Tr(l,k)$ useless rounds
in $[r1,r3-1]$.  If $r3 > r2$, it immediately follows that there
are at most $\Tr(l,k)$ useless rounds in $[r1,r2]$.  On the other
hand, if $r3 < r2$, since $l$ is still active at $r > r2$, it follows
that there are no useless rounds in $[r3,r2]$.  Hence, we again get
that there are at most $\Tr(l,k)$ useless rounds in $[r1,r2]$.
The lemma follows.
\end{proof}

The next lemma shows that in every execution of Protocol \B, by the
time a process becomes active, all lower numbered processes have retired.
 
\begin{lemma}
\label{properorderLemma}
In every execution of Protocol $\B$,
\begin{itemize}
\item[(a)]
before the round $r$ in which
process $i$ becomes preactive, all processes in groups
$<g_i$ have retired;
\item[(b)]
before the round $r$ in which
process $i$ becomes active, all processes $<i$ have retired.
\end{itemize}
\end{lemma}
 
\begin{proof}
Fix an execution $e_{{\cal B}}$ of Protocol $\B$.
The proof proceeds by induction on the round $r$.
The base case of $r=0$ holds trivially since only process 0 is active then.
Assume the claim for $<r$, and we will show it for $r$.
If $i=0$, the claim holds trivially.  Thus, we can assume $i>0$.
Suppose that the last ordinary message that $i$ received before round $r$
came from $k$, and was received at round $r1$.  (Note that there must
have been such an ordinary message, given our assumption that
process 0 sent an ordinary message to all the processes before the
execution begins.)

We first prove part (a).
Assume, by way of contradiction, that
some process $l$ with $g_l < g_i$ does not retire by round $r$.

Since, by assumption, $r1$ was the latest round $< r$ at which $i$
received an ordinary message, to complete the proof it is enough to show that if
$l$ does not retire before round $r$, $i$ must have received an ordinary message
at some round $r''$ with $r1  < r'' < r$.
In fact, we plan to show that $i$ must have received a message
in the interval $(r1,r)$ from some process in $\ac(l,r)$.  To do
this, we plan to use Lemmas~\ref{maxLemma} and~\ref{minLemma}.
Notice that both of these lemmas require $l$ to be active.  In
fact, we can assume without loss of generality that $l$ is active
at some round $r3 \ge r$ of
$e_{{\cal B}}$, and that $\ac(l,r) = \ac(l,r3)$.  If not, we can just
consider the execution $e'_{{\cal B}}$ which is identical to
$e_{{\cal B}}$ up
to round $r$, after which all processes other than $l$ crash.
It is clear that eventually $l$ becomes active in $e_{{\cal B}}$, with
the same activation chain it has in round $r$.  Moreover, if $i$
receives an ordinary message in the interval $(r1,r)$ in
$e'_{{\cal B}}$, then
it must also receive the same message in $e_{{\cal B}}$, since the two
executions agree up to round $r$.
 
Since $k$ becomes active at some
round prior to $r$,
the inductive hypothesis on part (b) of the lemma implies that all
processes $\le k$ have retired by round $r1 < r$. Thus without loss of
generality $l \ge k$. We consider two cases: (i) $k$ is in
$\ac(l,r)$; (ii) $k$ is not in $\ac(l,r)$.
 
In case (i), since $k$ is in $l$'s activation chain and is active at round $r1$,
by the inductive hypothesis, it must be the current process in
$\ac(l,r)$ at round $r1$.
Applying Lemma~\ref{minLemma} to
$\ac(l,r3) = \ac(l,r)$, we get
$$T^{r-1}_{r1+1}(l) \ge  (r-1)-(r1+1)-\Tr(l, k)+1.$$
 
By definition, $i$ becomes preactive in round $r=r1+\DDB(i,k)$, and
hence $r-r1=\DDB(i, k)$. Substituting this into the above inequality we get
$$T^{r-1}_{r1+1}(l) \ge  \DDB(i, k)-\Tr(l, k)-1.$$
 
Since $g_i > g_l$, Lemma~\ref{addLemma} implies that $\DDB(i, k)-\Tr(l, k) =
\DDB(i, l)$, and substituting this fact in the above inequality we get
\begin{eqnarray*}
T^{r-1}_{r1+1}(l)
&\ge&
\DDB(i, l)-1\\
&=&
\mdg(l)+(g_i-g_l-1)\mdg(0)-1\\
&=&
(n/\sqrt{t}+3\sqrt{t}+(\sqrt{t}-\overl-1)\mpd+1)+(g_i-g_l-1)\mdg(0)-1\\
&\ge& n/\sqrt{t}+3 \sqrt{t}.
\end{eqnarray*}
 
Thus part (b) of
Lemma~\ref{maxLemma}
implies that $i$ must have received an
ordinary message at some round in the interval $(r1, r)$, contradicting the
assumption that it does not, and the claim follows.
 
In case (ii), let $k'$ be the greatest process $<k$ in $l$'s activation chain,
and let $j$ be the smallest process $>k$ in $l$'s activation chain.
Suppose $j$ gets its last message before becoming active from $k'$ at round
$r0$. (Note that this means that the last message received by $j$ before
becoming active came at $r0$.)
Since the inductive hypothesis on part (b) implies that $k'$ must retire
before $k$ becomes active, and since $k$ must become active at least one
round before it sent a message to $i$
(since by assumption $g_k \le g_l < g_i$ and process
$k$ checkpoints to its own group before it sends a message to another
group),
we have $r0 < r1-1$. Furthermore, since the processes succeeding $k'$
in $l$'s chain are greater than $k$, the same inductive hypothesis implies that
these processes can become active only after process $k$ retires, and hence
after round $r1$. Since by definition, any message received by $i$ after round
$r0$ from $l$'s chain must be sent by one of the processes succeeding $k'$ in
the chain, it follows that if $i$ receives a message from $l$'s chain after
round $r0$, this message is sent after round $r1$.
 
To complete the proof we show that $i$ must have received
some message from $l$'s chain at some round in the interval $(r0, r)$, and hence
in the interval $(r1, r)$, contradicting the assumption that it does not.
As argued above, to show this it is enough to show that $T^{r-1}_{r0+1}(l) \ge
n/\sqrt{t}+3 \sqrt{t}.$  Applying Lemma~\ref{minLemma} to $l$'s activation chain
we get
$$T^{r-1}_{r0+1}(l) \ge r-1-(r0+1)-\Tr(l, k')+1.$$
 
To bound $T^{r-1}_{r0+1}(l)$, we need to bound $r-r0$. To do this we will compute
two terms: (a) $r-r1$; and (b) $r1-r0$.
The first term is equal to $\DDB(i, k)$ as argued above.
To compute the second term, we first show: (1) $g_j>g_k$, and (2)
$g_{k'}=g_k$.
 
For (1), clearly $g_j \ge g_k$, since $j > k$.
If $g_j=g_k$, $j$ must have received a message from $k$ at round $r1-1$ before
$k$ sent a message to $i$
(since by assumption $g_k \le g_l < g_i$ and process
$k$ checkpoints to its own group just before it sends a message
to another group).
As we have observed, $r1-1 > r0$, so this contradicts the
assumption that the last
message received by $j$ before becoming active came at $r0$.  Thus $g_j >
g_k$.
\remove{orli14:
We claim that no process other than $k$ sends an ordinary message to $j$ after
round $r1-1$ and before $j$ becomes active. This follows because on one hand,
since $j$ is the first process $>k$ in $l$'s activation chain, the last ordinary
message received by $j$ before it becomes active must be from a process $\le k$;
on the other hand, the inductive hypothesis on part (b) implies that all
processes $<k$ have retired by round $r1$. Since $k$ is the last process from
which $j$ receives a message before becoming active, $k$ must precede $j$ in
$l$'s activation chain, contradicting the assumption that $k$ is not in $l$'s
activation chain.}
 
For (2), clearly $g_{k'} \le g_k$. If $g_{k'} < g_k$, this means that $j$ did
not receive a message from a process in $g_k$ before becoming active
(because if it did, then by the inductive hypothesis on part (b) we have
that this message arrives after $k'$ retires and hence after round $r0$). But
since $g_j \le g_l < g_i$, and $k$ sent a message to $i$ at $r1$, some process in
$g_k$ must have sent a message to $j$ before round $r1$, and hence before $j$
becomes active. This gives us the desired contradiction.
 
\remove{orli14:
For (2), observe that $g_j \le g_l < g_i$. Hence $j$ must receive
a message from some process in $g_k$ before $i$ receives a message from $k$.
Let $k''$ be the last process in $g_k$ from which $j$ receives a message before
$j$ becomes active. To complete the proof it is enough to show that $k''$
immediately precedes $j$ in $l$'s activation chain. But this follows because, on
one hand, since $j$ is the first process $>k$ in $l$'s activation chain, the
last ordinary message received by $j$ before it becomes active must be from a
process in some group $\le g_k$; on the other hand, the inductive hypothesis on
part (b) implies that all processes in groups $<g_k$ have retired by the time
$k''$ sends a message. Therefore, the last message received by $j$ must be from
a process from $g_k$, and hence from $k''$, and hence $k''$ must immediately
precede $j$ in $l$'s activation chain.
}
 
To complete the proof of case (ii) we use the following claim:
 
\begin{claim}
\label{nolessClaim}
Every process $k''$ with $k' < k'' \le k$ that becomes active does so
no earlier than round $r0+(\bar{k''}-\bar{k'})\mpd-1$.
\end{claim}
 
\begin{proof}
\remove{orli12: replaced the next sentence with induction.
Assume the claim does not hold and we will get a contradiction. Let $k1$ be the
smallest process to violate the claim.
}
We proceed by induction. Assume $k' < k1 \le k$ and the claim holds for all
$k''$ with $k' < k'' < k1$. We prove it for $k1$.
 
We first show that the last
ordinary message that any process $\ge k1$ in $g_k$ receives from any process
$k2$ with $k' \le k2 < k1$ is sent no earlier than round
$r0+(\bar{k2}-\bar{k'})\mpd-1$. Observe that since $k'$ is in $g_k$, so are $k1$
and $k2$. If $k2=k'$, the claim trivially follows since $k'$ must send a message
to its own group at round $r0-1$ just before it sends a message to $g_j > g_k$.
Otherwise,
by the induction hypothesis we have that
$k2$ became active no earlier than round $r0+(\bar{k2}-\bar{k'})\mpd-1$, and the
claim again follows.
 
Let $k2$ be the last process from which $k1$ receives an ordinary message.
Observe that $k' \le k2 < k1$. (Because, as reasoned above, $k1$ has received a
message from $k'$, and hence the message sent from $k2$ was sent at or after the
time the message from $k'$; the inductive hypothesis on part (b)
therefore implies that $k2 \ge k'$.) It follows from the claim above that the
message from $k2$ was sent no earlier than round $r0+(\bar{k2}-\bar{k'})\mpd-1$.
In addition, the inductive hypothesis on part (b) implies that $k1$ becomes
active only after $k2$ retires, and hence only after receiving its message.
 
Now, assume that
$k1$ does not receive a \goahead~message. It then becomes
preactive $\mpd$ rounds after it receives the last ordinary message from $k2$
and then $k1$ starts sending \goahead~messages to lower numbered processes in its
group. Since, by assumption, $k1$ does not receive a message in response, it becomes
active $\Tr(k1, k2)=(\bar{k1}-\bar{k2)}\mpd$ rounds after receiving this last
message from $k2$, and hence no earlier than round
$r0+(\bar{k1}-\bar{k'})\mpd-1$.

Next assume $k1$ receives a \goahead~message. Let $k3$ be the process
sending this message. Let $k2$ be the last process from which $k3$ received an
ordinary message before sending the \goahead~message to $k1$. Since
$k3$ sends a \goahead~message to $k1$, it follows that $k2 <k1$.
Just as above, we can show
that $k2 \ge k'$, and hence that $k3$ received the
ordinary message from $k2$ no earlier than round $\ge
r0+(\bar{k2}-\bar{k'})\mpd-1$. Clearly, $k3$ sends the \goahead~message to $k1$
no earlier than $(\bar{k1}-\bar{k2})\mpd$ rounds after it receives its ordinary
message from $k2$, and the claim follows as above.  This completes
the proof of the inductive step.
\end{proof}
 
Now, to compute $r1 - r0$, observe that $r1$, the round in which $k$ sends a
message to $i$, is at least one round after $k$ becomes active (because $g_i >
g_k$ and $k$ first broadcasts to its own group), and hence
Claim~\ref{nolessClaim} immediately implies that $r1-r0 \ge
(\bar{k}-\bar{k'})\mpd$.  Thus we get that
\begin{eqnarray*}
T^{r-1}_{r0+1}
&\ge&
(r-r1)+(r1-r0)-\Tr(l, k')-1\\
&\ge&
\DDB(i, k)+(\overk-\bar{k'})\mpd-\Tr(l, k')-1\\
&=&
\mgd(k)+(g_i-g_k-1)\mgd(0)+(\overk-\bar{k'})\mpd-\Tr(l, k')-1\\
&=&
\mgd(k')+(g_i-g_k-1)\mgd(0)-\Tr(l, k')-1\\
&=&
\mgd(k')+(g_i-g_{k'}-1)\mgd(0)-\Tr(l, k')-1\\
&=&
\DDB(i, k')-\Tr(l, k')-1.
\end{eqnarray*}
(The fourth inequality follows because $\mgd(k)+(\overk-\bar{k'})\mpd=\mgd(k')$,
and the fifth inequality follows because $g_k=g_{k'}$.)
 
Again, Lemma~\ref{addLemma} implies that $\DDB(i, k')-\Tr(l, k')= \DDB(i, l)$,
and hence
\begin{eqnarray*}
T^{r-1}_{r0+1}
&\ge&
\DDB(i, l)-1\\
&=&\mgd(l)+(g_i-g_l-1)\mgd(0)-1\\
&\ge& (n/\sqrt{t}+3\sqrt{t}+(\sqrt{t}-\overl-1)\mpd+1)-1\\
&\ge& n/\sqrt{t}+3\sqrt{t}.
\end{eqnarray*}
This completes the proof of the inductive step for part (a).
 
\medskip
 
For part (b), suppose by way of contradiction that
$i$ becomes active at round $r$ and
process $l <i$ has not retired by round $r$.
First assume $i$ does not receive a \goahead~message.
If $g_l < g_i$, we get an immediate contradiction using the inductive
step for part (a),
since $i$ becomes active at or after it becomes preactive.
Otherwise, recall that
$k$ is the last process from which $i$ receives an ordinary
message before becoming active, and this message is received at round
$r1$. If $k>l$, then since $k$ became active before round $r$, the inductive
hypothesis on part (b) implies that $l$ must have retired before $k$ became
active and hence before round $r$. If $k=l$, then $i$ becomes preactive only
after $\mpd-1=n/t+1$ additional rounds in which it does not
hear from $l$. We claim that $l$ must have retired by that time.
Because otherwise, in this period $l$ would have either performed a subchunk and
informed its group, or would have checkpointed a subchunk to a group $g \ne g_l$
and informed its  group about the checkpoint. Since $g_i=g_l$, in both cases,
$i$ must have heard from $l$. Finally, if $l>k$, then before $i$ becomes active
it sends a \goahead~message to $l$ and waits for a message from $l$ for $\mpd-1$
additional rounds. Exactly as above, it follows again that since $i$ does not
receive any message from $l$, $l$ must have retired.
 
Next assume $i$ does get a \goahead~message before becoming active. However,
the same reasoning as above shows that by the time a process sends a
\goahead~message to process $i$, all processes $<i$ have retired, and we are
done.
\end{proof}

Finally we can show:
 
\begin{theorem}
\label{protocolBTheorem}
In every execution of Protocol \B,
\begin{itemize}
\item[(a)] at most $3n$ units of work are performed in total
by the processes,
\item[(b)] at most $10t \sqrt{t}$ messages are sent,
\item[(c)] by round $3n+8t$ all processes have retired.
\end{itemize}
\end{theorem}
 
\begin{proof}
Parts (a) and (b) were argued in the beginning of Section~\ref{section:
Bproof}.
 
For part (c), let $i$ be the last process that is active and
consider its activation chain. We want to find the last round $r2$
in which $i$ is active. It follows from Lemma~\ref{maxLemma}
that the maximal number of useful rounds performed by any chain is
$n+3t$. Therefore, applying Lemma~\ref{minLemma} with $k=0$
we get that
$$n + 3t \ge T^{r2}_1(i) \ge r2 - 1 - \Tr(i,0) + 1.$$
\remove{orli12: multiplying by \sqrt{t} did not give me the same complexity
since i culd not get rid of the plus 1.}
Thus
\begin{eqnarray*}
r2
&\le&
n+3t+\Tr(i, 0) \\
&\le&
n+3t+\Tr(t-1, 0) \\
&=&
n+3t+(\sqrt{t}-1)\mgd(0)+(\sqrt{t}-1)\mpd \\
&=&
n+3t+(\sqrt{t}-1)(n/\sqrt{t}+3\sqrt{t}+(\sqrt{t}-1)(n/t+2)+1)+(\sqrt{t}-1)(n/t+2)
\\
&\le&
n+3t+(\sqrt{t}-1)(n/\sqrt{t}+3\sqrt{t}+\sqrt{t}(n/t+2)+1)\\
&=&
n+3t+(\sqrt{t}-1)(n/\sqrt{t}+3 \sqrt{t}+n/\sqrt{t}+2 \sqrt{t}+1)\\
&\le&
n+3t+\sqrt{t}(2 n/\sqrt{t}+5 \sqrt{t})\\
&\le&
3n+8t.
\end{eqnarray*}
\end{proof}

\section{An Algorithm with Effort $O(n + t \log t)$}
 
In this section we prove that the effort  of
$O(n+t\sqrt{t})$ obtained by
the previous protocols is not optimal, even for work-optimal protocols.
We construct another
work-optimal
algorithm, Protocol~${\cal C}$,
that requires only $O(n+t \log t)$ messages (and a variant that requires
only $O(t \log t)$ messages),
yielding a total effort of $O(n+t \log{t})$.
As is the case with Protocols ${\cal A}$~and~${\cal B}$,
at most one process is active at any given time.
However, in Protocol ${\cal C}$ it is not the case that there
is a predetermined order in which the processes become active.
Rather, when an active process fails, we want the process that is
currently {\em most knowledgeable} to become the new active process.
As we shall see, which process is most knowledgeable after an active
process $i$ fails depends on how many units of work $i$
performed before failing.  As a consequence, there is no obvious variant
of Protocol ${\cal C}$ that works in the model
with asynchronous processes and a failure-detector.

Roughly speaking, Protocol ${\cal C}$
strives to ``spread out'' as uniformly as possible the
knowledge of work that has been performed and the processes that have
crashed.  Thus, each time the active process, say $i$, performs a new
unit of work or detects a failure, $i$ tells this to the process $j$ it
currently considers least knowledgeable.
Then process $j$ becomes as knowledgeable as $i$, so
after performing the next unit of work
(or detecting another failure), $i$ tells
the process it now considers least knowledgeable about this new fact.
 
The most na\"{\i}ve implementation of this idea is the following:
Process 0 begins by performing unit 1 of work and reporting this to
process 1.  It then performs unit 2 and reports units 1 and 2 to process
2, and so on, telling process $i$ mod $t$ about units 1 through $i$.
Note that at all times, every process
knows about all but at most the last
$t$ units of work to be performed.
 
If process 0 crashes, we want the most
knowledgeable alive process---the one
that knows about the most units of work that have been done---to become
active.  (If no process alive knows about any work, then we want the
highest numbered alive process to become active.)
It can be shown that
this can be arranged by setting appropriate deadlines.
Moreover, the deadlines are chosen so that
at most one process is active at a given time.  The most knowledgeable
process then continues to perform work, always informing the least
knowledgeable process.
 
The problem with this na\"{\i}ve algorithm is that it requires $O(n+t^2)$
work and $O(n+t^2)$ messages in the worst case.  For example, suppose
that process 0 performs the first $t-1$ units of work, so that the
last process to be informed is process $t-1$, and then crashes.
In addition, $t/2
+ 1, \ldots, t-1$ crash.
Eventually process $t/2$, the most
knowledgeable non-retired process, will become
active. However, process $t/2$ has no way of knowing whether process 0
crashed just after informing it about work unit $t/2$, or process 0
continued to work, informing later processes (who must have crashed, for
otherwise they would have become active before process $t/2$).  Thus,
process $t/2$ repeats work units $t/2 +1, \ldots, t-1$, again informing
(retired)
processes $t/2 +1, \ldots t-1$.   Suppose process $t/2$ crashes after
performing work unit $t-1$ and informing process $t-1$.
Then process $t/2
-1$ becomes active, and again repeats this work.  If each process $t/2
-1, t/2 -2, \ldots, 1$, crashes after repeating work units $t/2 + 1,
\ldots, t-1$, then $O(t^2)$ work is done, and $O(t^2)$ messages are sent.
(A slight variant of this example gives  a scenario in which
$O(n+t^2)$ work is done, and $O(n+t^2)$ messages are sent.)
 
\remove{orli: added instead the paragraph below. To prevent this
situation, in our algorithm, a process does failure detection before proceeding
on the work.  Essentially, we treat failure detection as another type of work.
One level of failure detection turns out not to be enough to prevent $O(t^2)$
messages.  In order to get down to $O(t \log t)$, we must do failure detection
recursively;
before doing the $k$th level of failure detection, a process must
complete the $(k+1)$st level of failure detection.  This point
will become clearer when we present our Protocol ${\cal C}$ formally.
}
 
To prevent this situation, a process
performs failure detection before proceeding with the work.
The key idea here is
that we treat failure detection as another type of work.
This allows us to use our algorithm recursively for failure detection.
\remove{
This approach encounters two main difficulties.
The first difficulty is that
finding an efficient failure detection algorithm
reduces then to our original
problem of finding an efficient algorithm for
performing work in the presence of
failures. This difficulty is
overcome by observing that the algorithm required
for failure detection may have
slightly higher relative complexity than the original one required for
performing the real work. This follows from the fact that the size of the
failure detection `work'
is only $t$, since at most $t$ processes may become
faulty throughout the execution. The second, more serious difficulty}
Specifically, fault-detection is accomplished by polling a process
and waiting for a response or a timeout.
The difficulty
encountered by our approach is that, in contrast to the real work,
the set of faulty processes
is dynamic, so it is not obvious how these processes
can be detected without sending (wasteful) polling
messages to {\em nonfaulty\/} processes.
In fact, in our algorithm we do not attempt to detect all the faulty
processes, only enough
to ensure that not too much work is wasted by reporting work to faulty
processes.
 
\begin{figure*}[htb]
\begin{tabbing}
XX \= XXXX \= XXXX \= XXXX \= XXXX \= XXXX \= XXXX
\kill
\\
1. \>$h \gets \log t$;\\
2. \>{\bf while} $ h > 0$ {\bf do} \\
3. \>\> $\found \gets \false;$\\
4. \>\> {\bf while} $\neg \found$ {\bf do}\\
5. \>\>\> Send ``Are you alive?'' to $\point_i[G^i_h]$;\\
6. \>\>\> {\bf if} no response\\
7. \>\>\> {\bf then} \> add $\point_i[G^i_h]$ to $F_i$; \\
8. \>\>\>\> {\bf if} $h \ne \log t$\\
9. \>\>\>\>{\bf then} \>send ordinary message to $point_i[G^i_{h+1}]$;\\
10. \>\>\>\>\>$\round_i[G^i_{h+1}] \gets \curr$;\\
11. \>\>\>\>\> $\point_i[G^i_{h+1}] \gets \succ(\point_i[G^i_{h+1}])$;\\
12. \>\>\>\> {\bf if} $G^i_h - F_i \ne \{i\}$\\
13. \>\>\>\>{\bf then}
\>$\point_i[G^i_{h}] \gets \succ(\point_i[G^i_{h}])$;\\
14. \>\>\>\>{\bf else} \>  $\found \gets \true$;\\
15. \>\>\> {\bf else} \> ({\it i.e.}, response
              received) $\found \gets \true$;\\
16. \>\> $h \gets h-1$;\\
\\
 
\>\{Process level 0 (real work):\}\\
17. \> {\bf while} $\point_i[G_0] \le n$ {\bf do}\\
18. \>\> Perform work unit $\point_i[G_0]$;\\
19. \>\> Send an ordinary message to $\point_i[G^i_1]$; \\
20. \>\> $\round_i[G^i_1] \gets \curr$; \\
21. \>\> $\point_i[G^i_1] \gets \succ(\point_i[G^i_1])$; \\
22. \>\> $\point_i[G_0] \gets \succ(\point_i[G_0])$
 
\end{tabbing}
\caption{Code for Active Process $i$ in Protocol $\C$}
\label{figure: code for an active process}
\end{figure*}

\subsection{Description of the Algorithm}
 
For ease of exposition we assume $t$ is a power of 2.
Again, the processes are numbered $0$ through $t-1$, and the units
of work are numbered 1 through $n$.
Although  our algorithm is recursive in
nature, it
can more easily be described when the recursion is unfolded.
Processing is divided into
$\log{t}$ levels, numbered 1
to $\log{t}$,
where level $\log{t}$ would have been the deepest
level of the recursion, had we presented the algorithm recursively.
In each level, the processes are partitioned into
groups as follows.
In level $h$, $1 \le h \le \log t$, there are $t/(2^{\log t -h + 1})$
groups of size $2^{\log t -h +1}$.
Thus,
in level $\log{t}$, there are $t/2$ groups of size $2$,
in level $\log t -1$ there are $t/4$ groups of size $4$,
and so on, until level 1, in which there is a single group
of size~$t$.
Let $s_h= 2^{\log t -h +1}$ denote the size of a group at level~$h$.
The first group of level $h$ contains processes
$0,1, \dots, s_h-1$, the next group contains processes
$s_h, s_h +1, \dots, 2s_h-1$, and so on.
Thus each group of level $h < \log t$ contains
two groups of level $h+1$.
Note that each process $i$ belongs to $\log t$ groups, exactly
one on each level.
We let $G^i_h$ denote the level~$h$ group of process~$i$.
 
Initially process 0 is active.  When process~$i$
becomes active, it performs fault-detection in its group at
every level, beginning with the highest level and working its
way down, leaving level $h$ as soon as it finds a non-faulty
process in $G^i_h$.
Once fault-detection has been completed on $G^i_1$, the set
of all processes, process $i$ begins to perform real work.
Thus, we sometimes refer to the actual work as $G_0$, or
level 0, and the
fault-detection on level $h$ as {\it work on level $h$}.
For each $1 \le h \le \log t$,
each time it performs a unit of work
on $G^i_{h-1}$, process~$i$ reports that work to some
process in $G^i_h$.
\remove{orli14: added brackets.}
(Observe that the above protocol requires at least $n$ messages. However, it
will later become clear that modifying this protocol so that
when a process performs work on $G_0$, it reports only each time it
completes $n/t$ units of work, will immediately give a work optimal
protocol that requires only $O(t \log t)$ messages.)
 
A {\em unit of
fault-detection} is performed by sending a special message
``Are you alive?'' to one process, and waiting for a reply
in the following round.
An {\it ordinary} message
informs a process at some level $h$,
$1 \le h \le \log t$, of a unit of (real or fault-detection) work
at level $h-1$.
As we shall see, an ordinary message also carries additional information.
These two are the only types of messages sent by an active process.
As before, a process that has crashed or terminated is said to be
{\em retired}.
An inactive non-retired process only sends responses to
``Are you alive?'' messages.
 
Each process $i$ maintains a list $F_i$ of processes
known by $i$ to be retired.
It also maintains an array of pointers,
$\point_i$, indexed by group name.
Intuitively, $\point_i[G_0]$ is the successor of the last unit of work
known by $i$ to have been performed
(and therefore this
is where $i$ will start
doing work when it becomes active).  For
$h \ge 1$, $\point_i[G^j_h]$ contains
the successor (according to the cyclic order in $G^j_h$,
which we define precisely below)
of the last
process in $G^j_h$ known by $i$ to have received an
ordinary message from a process in $G^j_h$ that was
performing (real or fault-detection) work on $G^j_{h-1}$.
We call $\point_i[G^j_h]$ {\em process $i$'s pointer into
$G^j_h$}.
Process $i$'s moves are governed entirely by
the round number, $F_i$, and pointers into its own
groups (i.e., pointers into groups $G^i_h$).
Associated with each pointer $\point_i[G]$
is a round number, $\round_i[G]$,
indicating
the round at which the last message known to be sent was sent (or, in
the case of $G_0$, when the last unit of work known to be done was
done).
Initially, $\point_i[G_0] = 1$, $\point_i[G^j_h]$ is the
lowest-numbered process in $G^j_h-\{i\}$, and $\round_i[G_0] =
\round_i[G^j_h] = 0$.
We occasionally use $\round_i[G](r)$ to denote
the value of $\round_i[G]$ at the beginning of round $r$; we
similarly use $F_i(r)$ and $\point_i[G](r)$.
 
\begin{sloppypar}
The triple
$(F_i, \pointer_i,\round_i)$ is the {\em view} of process~$i$.
We also define the {\em reduced view} of
process~$i$ to be $\pointer_i[G_0]-1 + |F_i|$;
thus, $i$'s reduced view is the sum of the number of units of work known
by $i$ to be done and the number of processes known by $i$ to be faulty.
A process includes its view whenever it sends an ordinary
message. When process $i$ receives an ordinary message, it updates
its view in light of the new information received.
Note that process $i$
may receive information about one of its
own groups from a process not in that group.
Similarly, it may pass to another process information
about a group in which the other process is a member
but to which $i$ does not belong.
\end{sloppypar}

Let $G^i_h$ be any group as described above, where
the process numbers range from $x$ to $y = x + |G^i_h|-1$.
There is a natural fixed cyclic order on the group,
which we call the {\em cyclic order}.
Process $i$
sends
messages to members of
$G^i_h$ in
{\it increasing order}.  By this we mean
according to the cyclic order but
skipping itself
and all processes in $F_i$.
Let $j\not= i$ be in $G^i_h$.
Then $j$'s {\em $i$-successor
in $G^i_h$},
is $j$'s nearest successor in the cyclic ordering
that is not in $\{i\} \cup F_i$.
We omit the $i$ in ``$i$-successor,'' as well
as the name of the group in which the successor is
to be determined, when these are clear from the
context.

\remove{
For ease of description, we assume each group $G^i_h$, $h > 0$,
includes two
special members, $\bot$ and $\top$.
The $i$-successor of $\bot$ is the smallest-numbered
process in the group.
The arrays $\pointer_i$ are all
initialized to $\bot$, except $\point_i[G_0]$, for all processes~$i$,
which is initialized to~0.
The $i$-successor of $\top$ is itself.
If
$G^i_h \setminus \{i\} \subseteq F_i$, so that all processes in $G^i_h$
other than $i$ are known by $i$ to be faulty, then
$\succ(\point_i[G^i_h])
= \top$, regardless of the value of $\point_i[G^i_h]$.
Sending a message to $\top$ is a no-op.}
 
When process $i$ first becomes active it searches
for other non-retired processes as follows.
For each level $h$, starting with $\log t$ and going down to 1,
process $i$
polls group $G^i_h$, starting with
$\pointer_i[G^i_h]$, by sending an ``Are you alive?'' message.
If no answer is received,
it adds this process to $F_i$.
\remove{orli: added the beginning of the sentence}
If $h < \log t$, process $i$ sends an ordinary message reporting
this newly detected failure to
$\pointer_i[G^i_{h+1}]$,
sets $\pointer_i[G^i_{h+1}]$ to its $i$-successor
in $G^i_{h+1}$,
and
sets $\round_i[G^i_{h+1}]$ to the current round number.
Process $i$ repeats these steps until an
answer is received or $G^i_h \setminus \{i\} \subseteq F_i$.
It then enters level $h-1$, and repeats the process.
Note that if no reply was received, then
although the
pointer into $G^i_h$ does not change, the successor
in $G^i_h$ of $\pointer_i[G^i_h]$ does change, because
the successor function takes into account $F_i$, which
has changed.
 
Level 0 is handled similarly to levels 1 through $\log t -1$, but
the process performs real work
instead of polling,
and increases the work pointer after performing each unit of  work.
If $\point_i[G_0] = n+1$ then process $i$ halts, since
in this case all the work has been completed.
This completes the description of the behavior of an active
process.
The code for an active process
appears in Figure~\ref{figure: code for an active process}.

At any time in the execution of the algorithm, each inactive
non-retired process $i$ has a {\it deadline}.
We define $D(i,m)$ to be the number of rounds
that process $i$ waits from the round in which it first obtained
reduced view $m$ until it becomes active:
$$D(i,m) = \left\{ \begin{array} {ll}
            K(n+t-m)2^{n+t-1-m} & \mbox{if $m \ge 1$} \\
            K(t-i)  (n+t) 2^{n+t-1} & \mbox{otherwise.}
                   \end{array}
           \right. $$
\noindent
where $K = 5t + 2\log t$.
As we show below (Lemma~\ref{upperbound}), $K$ is an upper bound
on the number of rounds that any process needs
to wait before first
hearing from the active process.  (More formally, if $j$
becomes active at round $r$ and is still active $K$ rounds later,
then by the beginning of round $r+K$, all processes that are not
retired will have received a message from $j$.)
All our arguments below
work without change if we replace $K$ by any other bound on the number
of rounds that a process needs to wait before first hearing from the
active process.  This observation will be useful later,
when we consider a
slight modification of Protocol ${\cal C}$.

If process $i$ receives no message by the end of $D(i,0) -1$, then
it becomes active at the beginning of round $D(i,0)$.
Otherwise, if at round $r$ it receives a message based on which
it obtains a reduced view of $m$, and if it receives no
further messages by the end of
round $r+D(i,m)-1$, it becomes
active at the beginning of round $r+D(i,m)$.
This completes the description of the algorithm.
 
\subsection{Analysis and Proof of Correctness}

\begin{lemma}
\label{doneLemma}
In every execution of Protocol $\C$ in which there are no more
than $t-1$ failures, the work is completed.
\end{lemma}
 
\begin{proof}
By assumption, one of the processes is correct, say $i$.
At some point process~$i$ will become active, since once every
other process has retired process $i$ will not extend its deadline.
It is straightforward from inspection of the algorithm that
at any time during the execution of the algorithm
$\pointer_i[G_0] = w$ if and only if the first
$w-1$ units of work have been performed, and that when it
becomes active, process~$i$
performs all units of work from $\point_i[G_0]$ through~$n$.
\end{proof}

The next lemma shows that our choice of $K$ has the properties mentioned
above.
\begin{lemma}\label{upperbound}
If $j$ is active at round $r$, and is not retired by round
$r+5t + 2\log t$, then all processes
that are not retired will receive a message from $j$ before the
beginning of $r+ 5t + 2\log t$.
\end{lemma}
 
\begin{proof}
It is immediate from the description of the algorithm that all
\remove{orli11: added nonretired}
nonretired processes have received a message from
$j$ by the time it has performed $t$ units of work (at level $G_0$) after round
$r$.  Thus, we compute an upper bound on the time
it takes for $j$ to perform $t$ units of work starting at round $r$.  In
the worst case, $j$ has just become active at the beginning of round
$r$, and must do failure detection before  reaching level
$G_0$ and doing work.
While doing this failure detection, $j$ sends ``are
you alive?'' messages to at most $t+\log t$ processes (the extra $\log t$
is due to the fact that at each level, it may send one ``are you
alive?'' message to a process that is alive, but crashes later while
$j$ is doing failure detection on a larger group).
After discovering a failure, process
$j$ sends an ordinary message; thus, it sends at most $t$ ordinary
messages.  Each message sent takes up one round; in addition, process
$j$ waits one round for a response after each ``are you alive?''
message.  This means that $j$ spends at most
\remove{orli11: replaced log t by 2log t}
$3t + 2\log t$ rounds
in levels $G^j_{\log t}, \ldots , G^j_1$.  Clearly, $j$ spends
\remove{orli11: added `\le'}
$\le 2t-1$ rounds working at level $G_0$ in the course
of doing $t$ units of work
(since it sends an ordinary message between each unit of work).
The required bound follows.
\end{proof}

If $i$ received its last ordinary
message from $j$ at round $r$, we call
other processes that received an ordinary message from $j$ after $i$ did
{\em
first-generation processes}
(implicitly, with respect to $i$, $j$, and $r$).
If $i$ did not yet receive any
ordinary messages,
then the first-generation processes (with respect to $i$ and $r$) are
those that received an ordinary
message from a process with a number greater
than $i$.
We define $k$th generation processes inductively.
If we have defined $k$th generation, then the $(k+1)$st
generation are those
processes that receive an ordinary
message from a $k$th generation process.
The {\em rank} of a process is the highest generation that it is in.

\begin{lemma}
\label{lemma: reduced view m+k}
Let $i$ receive its last ordinary message from $j$ at round~$r$,
let $m$ be the reduced view of $i$ after receiving  this message,
and let $\ell$ be a $k$th rank process with respect to $i$, $j$,
and $r$.
Then, after $\ell$ receives its last ordinary message,
its reduced view is at least $m+k$.
\end{lemma}
 
\begin{proof}
The proof is
an easy induction on $k$, since when a $k$th rank process
becomes active, it knows about everything its parent knew when it
became active, and at least one more piece of work or failure.
\end{proof}

We say process $i$  {\it knows more} than
process $j$ at round $r$ if $F_i(r) \supseteq F_j(r)$ and
for all groups $G$,
$\round_i[G](r) \ge \round_j[G](r)$.
Note that if equality
holds everywhere then intuitively the two processes
are equally knowledgeable.
We first show that our algorithm
has the property that for any two
inactive non-retired processes, one of them is
more knowledgeable than the other, unless they both
know nothing; that is, the knowledge of two
non-retired processes is never incomparable.
This is important so that the ``most knowledgeable''
process is well-defined.
Moreover, the knowledge can be quantified by the reduced view.
Process $i$ knows more than inactive
process $j$ if and only if the reduced view of $i$
is greater than the reduced view of~$j$.
Finally,
the algorithm also ensures that the active process is at least
as knowledgeable as any inactive non-retired process.

\begin{lemma}
\label{totalorderLemma}
For every round $r$ of the execution the following hold:
\begin{itemize}
\item[(a)]
If process $i$ received an ordinary message from process~$ j$
at round $r' < r$,
and $i$ is inactive and has not retired by the beginning of
round~$r$, then at the beginning of round $r$, no
processes other than $j$ and
processes in the $k$th generation with respect to $i$, $j$, and $r'$,
for some $k \ge 1$, know as much as~$i$.
\item[(b)]
Suppose process $i$ received its last ordinary message at round $r'$ (if
$i$ has received no ordinary messages then $r'=0$), and
$m$ is $i$'s reduced view after receiving this message.
If $i$ is not retired at the beginning of round $r=r'+D(i,m)$, and it
receives no further ordinary messages before the beginning
of round $r$, then at the beginning of round~$r$ no
non-retired process knows more than~$i$.
\item[(c)]
At the beginning of round~$r$,
there is an asymmetric total order (``knows more than'')
on the non-zero knowledge of the non-retired
processes
\remove{orli11: added because needed for the proof of part d}
that did not become active before round $r$, and the active
process knows at least as much as the most knowledgeable among these processes.
Moreover, for any two non-retired processes $i$ and $j$,
$i$ knows more than $j$ if and only if the reduced view of
$i$ is greater than the reduced view of $j$.
\item[(d)]
At most one process is active in round~$r$.
\end{itemize}
\end{lemma}
 
\begin{proof}
The proof is by induction on $r$.
The base case, $r=1$, is straightforward.
Let $r>1$, and
assume that all parts of the lemma hold for smaller values of $r$.
We prove it for~$r$.
 
For part (a), observe that by the inductive hypothesis, (a) holds at the
beginning of round $r-1$.  If no process is active in round $r-1$, then
no process' knowledge changes, so (a) holds at the beginning of round
$r$ as well.  If process $j'$ is active in round $r-1$, then by part (c)
of the inductive hypothesis, $j'$ knows at least as much as $i$.  Thus,
by part (a), it must be the case that $j'$ is either $j$ or some
process in the $k$th generation with respect to $i$, $j$, and $r'$, for
some $k$
(since, by assumption, $i$ is not active at the beginning of round $r$).
The only process whose knowledge changes during round $r-1$ is one to
which $j'$ sends an ordinary message.
It is immediate from the definition that
this process must be in the $k$th generation with respect to $i$, $j$,
and $r$, for  some $k$.
 
For part (b), we must consider two cases: $r' > 0$ and $r' = 0$.
If $r' >0$, let $j$ be the process that
wrote to $i$ at $r'$.
By part~(a) we have that
only $j$ and processes in generation $k \ge 1$ with
respect to $i$, $j$, and $r'$ are as knowledgeable as $i$ at any
round in the interval $[r',r)$.
By part~(c), these can be the
only processes active in this interval.  Thus, it suffices to argue
that $j$ and all processes of generation $k \ge 1$ with respect
to $i$, $j$, and $r'$ are retired by the beginning of round~$r$.
Since a reduced view is at most $n+t-1$, the
highest rank a process could be in is $n+t-1$.
We now argue that by the beginning of round
$r'+D(i,m) > r'+(n+t-m)K + D(i,m+1) + \cdots + D(i,n+t-1)$
all processes of ranks 1 through $n+t-1$ have
retired.
More generally, we argue by induction on $k$
that for every $k$ with $1 \le k \le n+t-m-1$,
by the beginning of
round $s + (k+1)K + D(i,m+1) + \cdots + D(i,m+k)$, every process in
ranks 1 to $k$ has retired.
 
If $k=1$, note that since $i$ received an ordinary
message from~$j$ at round~$r'$, by Lemma~\ref{upperbound},
every rank~1 process receives
a message from~$j$ before round $r'+K$.
By Lemma~\ref{lemma: reduced view m+k},
the reduced view of any such process is at least $m+1$.
Since $i$ receives no message from $j$ by round $r'+K$, it must
be the case that $j$ has retired by round $r'+K$.  By definition, no
rank~1 process can receive any messages at any round in $[r'+K,r)$
(otherwise it would have a rank higher than 1).
Thus, any rank~1 process $i'$ became active before $r'+K+D(i',m+1)$,
so by definition of $K$ and the fact that
$D(i',m+1) = D(i,m+1)$,
$i$ would have heard from $i'$ before $r'+2K+D(i,m+1)$.  It is easy to
check that $r' + 2K +D(i,m+1) < r' + D(i,m) = r$.  Since $i$ did not
receive
any messages by the beginning of round $r$, $i'$ must have
retired by then.
 
In general, consider a rank $k+1$ process $i'$,
and assume inductively that every rank~$k$ or lower process
has retired by the beginning of round~$r$.
By definition of rank, $i'$ received an ordinary
message from a rank~$k$ process, and, since these
are all retired by round~$r$, $i'$ must have
received this message before round~$r$.
By the inductive hypothesis on $k$, $i'$
must have received its last ordinary message
by the beginning of round
$r' + (k+1)K + D(i,m+1) + \cdots + D(i,m+k)<r$  (again using the fact
that $D(i,m) = D(i',m)$ if $m > 0$).
By Lemma~\ref{lemma: reduced view m+k},
the reduced view of $i'$
when it received its last ordinary message before round~$r$
was at least $m+k+1$.  Thus, it must have
become active before round $r'+(k+1)K + D(i,m+1) + \cdots + D(i,m+k+1)$,
if it became active at all.  Since $i$
received no messages from $i'$, it follows
that $i'$ must have retired before round
$r'+(k+2)K + D(i,m+1) + \cdots + D(i,m+k+1) < r$.
This completes the induction on~$k$.

If $r'=0$
we need the fact that
$D(i,0) > (n+t)K + max_{j>i} \{D(j,0)\} + D(i,1) + \cdots + D(i,n+t-1)$,
which follows easily from the definitions.
We claim that, for every $k \ge 0$, by round
$(k+1)K + max_{j> i} \{ D(j,0) \} + D(i,1) + \cdots + D(i,k)$,
every process
in ranks 1 to $k$ has retired.
To see this, note that a rank~0 process $j'$ (one with a higher
number than $i$ that received no messages) must have
become active at round~$D(j',0)$, and therefore must
have retired by round $D(j',0)+K$.
Thus a level~1 process received its last message
by $\max_{j>i} \{D(j,0)\}+K$.
We now proceed as in the case $r' >0$.
 
To prove part~(c), observe that the result is immediate from the inductive
hypothesis applied to $r-1$ if there is no active process at the
beginning of round $r-1$ (for in that case, no process' reduced view
changes).  Otherwise, suppose that $j$ is active at the beginning of
round $r-1$.  If $j$ does not send an ordinary message in round $r-1$,
again the result follows immediately from the inductive hypothesis
(since no process' reduced view changes).  If $j$ does send an ordinary
message to, say, process $i$, it is immediate that $i$ and $j$ know more
at the beginning of round $r$ than any other non-retired process, and
that $i$'s reduced view is greater than that of any other
non-retired inactive process.
 
It remains to show part~(d). Observe that the result is immediate if no
process becomes active at round $r$.  Now suppose that process $i$ becomes
active at the beginning of round $r$.  We must show that no process that
was active prior to round $r$ is still active at the beginning of round
$r$, and that no process besides $i$ becomes active at round $r$.  Let
$r'$ be the last round
in which $i$ received a message (as usual, if $i$ received no messages
prior to round $r$, then we take $r' = 0$), and suppose that $m$ was
$i$'s reduced view at round $r'$.  Then we must
have $r = r' + D(i,m)$. {F}rom part~(b), it follows that no
non-retired process knows more than~$i$ at the beginning of
round~$r$.  {F}rom part~(c), it follows that any process that was
active in the interval $[r',r)$ must know more than~$i$.  This
shows that all processes that were active before round $r$ must
have retired by the beginning of round~$r$.  Suppose some other
process $i'$ becomes active at round $r$.
\remove{orli11: added}
We have just shown that $i'$
does not know more than $i$. {F}rom part (c) it follows therefore that $i'$ knows
less than $i$. Thus part (b) provides a contradiction to the assumption
that $i'$ becomes active at round $r$.
\remove{orli11: The details left to the reader are not clear.
We have just shown that $i'$ cannot know more than $i$; a symmetric argument
shows that $i$ cannot know more than $i'$.
{F}rom part~(a) of the induction
hypothesis, it follows that $i$ and $i'$ must have the same reduced view $m$.
Let $r''$ be the round in which $i'$ received its last message.  Thus, we must
have $r = r' + D(i,m) = r'' + D(i',m)$. We leave it to the reader to check that
this is  impossible unless $i = i'$.}
\end{proof}

\begin{lemma}
\label{lemma: run time}
The running time of the
algorithm is at most
$tK(n+t)2^{n+t}$ rounds.
\end{lemma}
 
\begin{proof}
If process $i$'s
reduced view is $m$ and it does not receive
a message within $D(i,m)$ steps, then it becomes active.
Each message that $i$
receives increases its reduced view.
Thus, $i$ becomes active in at most
$D(i,0) + \cdots + D(i,n+t-1)$ rounds.
Once it becomes active, arguments similar to those used in
Lemma~\ref{upperbound} show that it retires
in at most $2n+3t+2\log t$ rounds.  Thus, the running time of the
algorithm is at most $D(1,0) + \cdots + D(1,n+t-1) + 2n + 3t + 2 \log t
\le tK(n+t)2^{n+t}$ rounds.
\end{proof}
 
The next lemma shows that an active process $i$
does not send messages
to retired processes that, because they were more knowledgeable
than $i$,
should have become active before $i$ did.
These messages are avoided
because during fault detection $i$ discovers that these
processes have retired.

\begin{lemma}\label{monotonicLemma}
If process $i'$ gets an ordinary message at round $r'$ from a
process operating on group $G_{h-1}^{i'}$ and process $i$ is
active at the beginning of round $r > r'$ then:
\begin{itemize}
\item[(a)] if $\round_i[G_h^{i'}](r) \ge r'$, then all processes
in the interval $[i',
\point_i[G_h^{i'}](r))$ in the cyclic order on $G_h^{i'}$ are
either
retired by the beginning of round $r$ or receive an ordinary
message in the interval
$[r',\round_i[G_h^{i'}](r)]$ from a process operating on $G_{h-1}^{i'}$.
(If $i' = \point_i[G_h^{i'}](r)$, then all processes in $G_h^{i'}$ are
either retired by the beginning of round $r$ or receive a message
in the interval $[r',\round_i[G_h^{i'}](r)]$ from a process operating on
$G_{h-1}^{i'}$.)
Moreover, either $i$'s
knowledge at the beginning of round $r$ is greater than $i'$'s knowledge
at the end of $r'$, or $i' \in F_i(r)$.
\item[(b)] if $\round_i[G_h^{i'}](r) < r'$, then all processes in the
interval
$[\point_i[G_h^{i'}](r), i']$
in the cyclic order on $G_h^{i'}$ are either
retired by the beginning of round $r'$, or receive a message in the
interval
$(\round_i[G_h^{i'}](r), r']$ from a process operating on $G_{h-1}^{i'}$.
Moreover, all the processes in this interval are retired by the
beginning of round $r$, and if
$G^i_h = G^{i'}_h$, then all these processes will be in $F_i$ by the time
$i$ begins to operate on $G^i_{h-1}$.
\end{itemize}
\end{lemma}
 
\begin{proof}
We proceed by induction on $r$.  The case $r=1$ is vacuous.  Assume
that $r > 1$ and the result holds for $r-1$.
If $r' = r-1$, then it must be the case that $i'$ received its
message from $i$, $\round_i[G_h^{i'}](r)= r-1$, and $\point_i[G_h^{i'}]$
is the successor of $i'$ in the cyclic order on $G_h^{i'}$, as computed
by $i$ in round $r-1$.  It is easy to see that the result follows
immediately in this case, because all processes in the interval
$(i',\point_i[G_h^{i'}])$ must be retired.
 
Suppose $r' < r-1$.
If $i$ is also active at round $r-1$, then the result is
immediate from the inductive hypothesis unless $\round_i[G^{i'}_h]$
changes during round $r-1$.  The description of the
algorithm shows that $\round_i[G^{i'}_h]$ changes only if $G^i_h =
G^{i'}_h$ and $i$ is operating on group $G^i_{h-1}$, in which case
$\round_i[G^{i'}_h]$ is set to $r-1$ at the end of round $r-1$,
and $\point_i[G^{i'}_h](r)$ is the successor of $\point_i[G^{i'}_h](r-1)$
in the cyclic order on $G^{i'}_h$.
In this case it is easy to see
that the result follows from
the inductive hypothesis; we leave details to the reader.
 
Thus, we have reduced to the case that $i$ becomes active
at round $r$.
Let
$\round_i[G_h^{i'}](r) = r''$ and let $\point_i[G_h^{i'}](r) = i''$.
If $r'' \ge r'$, then it must be the
case that
$i$ received a message from $j$ at some earlier round $s$ such that
$\point_j[G_h^{i'}](s) = \point_i[G_h^{i'}](r)$ and
$\round_j[G_h^{i'}](s) = \round_i[G_h^{i'}](r)$.
Since we must have $r' \le r'' = \round_j[G_h^{i'}](s) \le s$,
\remove{orli11: replaced above `< s' by `\le s'}
the result now
follows from the induction hypothesis (using $j$ and $s$ instead of
$i$ and $r$).
 
It remains only to consider the case
$r'' < r'$.  Let $j' \in G_h^{i'}$ be the process that sent the ordinary
message
to process $i'$ at round $r'$, and suppose that $j'$ became active
at the beginning of round $s'$.
We claim that we have the following chain of inequalities:
$r'' \le \round_{j'}
[G^{i'}_h](s') <s' < r' < r$.  Every inequality in this
chain is immediate from our assumptions except the first one.
Suppose that
$\round_{j'}[G^{i'}_h](s') < r''$.
{F}rom
Lemma~\ref{totalorderLemma}, it follows that $\round_k[G^{i'}_h](s') <
r''$ for all processes $k$ not retired by round $s'$.  This means that
no process not retired at $s'$ knows that a message was sent at round
$r''$.  But at round $r > s'$, process $i$ knows this fact (since, by
assumption $\round_i[G^{i'}_h](r) = r''$). This is impossible.  Thus, we
must have $\round_{j'}[G^{i'}_h](s') \ge r''$.
Note that $\point_{j'}[G^{i'}_h](r') = i'$, by assumption.  Thus,
by the inductive hypothesis, all processes in the cyclic order on
$G^{i'}_{h}$ in the interval $[i'',i')$ are either retired by the
beginning of round $s'$ or receive an ordinary message in the
interval $[r'',\round_{j'}[G^{i'}_h](s')]$ from a process operating on
$G_{h-1}^{i'}$.  Since we also know
that $i'$ receives a message at round $r'$ from a process operating
on $G_{h-1}^{i'}$, this proves the
first half of part (b).
Since, by Lemma~\ref{totalorderLemma}, all processes not
retired by round $r$
must be less knowledgeable than $i$
at the beginning of round $r$, it follows from
Lemma~\ref{totalorderLemma} that all the processes in the interval
$[i'',i']$ in the cyclic order have in fact
retired by round $r$.  {F}rom the description of the algorithm,
it follows that $i$ will detect this fact before it starts operating
on $G^i_{h-1}$.
\end{proof}

Observe that the algorithm treats `are you alive?' messages as real
work. Therefore,
we refer to these
messages as work unless stated otherwise. On the other hand, the
ordinary messages are still referred to as messages.
 
Using Lemma~\ref{monotonicLemma}, we can show that indeed effort is
not wasted:
\begin{lemma}
\label{wasteCorollary}
At most
$|G^i_h|+|G^i_{h-1}|$ units of work
are done and reported to $G^i_h$ by
group $G^i_h$ when operating on group $G^i_{h-1}$.
\end{lemma}

\begin{proof}
Given $i$, $h$, and an execution $e$ of Protocol ${\cal C}$, we consider
the sequence of triples
$(x,y,z)$, with one triple in the sequence for
every time a process $x \in G^i_h$ sends an ordinary message reporting a
unit of work $y \in G^i_{h-1}$ to a process $z \in G^i_h$, listed
in the order that the work was performed.
We must show that the length of this sequence is no greater than
$|G^i_{h-1}|+| G^i_h|$.
 
We say that a triple $(x,y,z)$ is
{\em repeated\/} in this sequence if there is a triple $(x',y,z')$ later
in the sequence where the same work unit $y$ is performed.  Clearly
there are at most $|G^i_{h-1}|$ nonrepeated triples in the sequence, so
it suffices to show that there are at most $|G^i_h|$ repeated triples.
To show this, it suffices to show that the third components of repeated
triples (denoting which process was informed about the unit of work)
are distinct.  Suppose, by way of contradiction, that there are two
repeated triples $(x_1,y_1,z_1)$ and $(x_2,y_2,z_1)$ with the same third
component.  Suppose that $x_1$ informed $z_1$ about $y_1$ in round $r'$,
and $x_2$ informed $z_1$ about $y_2$ in round $r''$.
Without loss of generality, we can assume that $r' < r''$.
Since
$(x_1,y_1,z_1)$ is a repeated triple, there is a triple $(x_3,y_1,z_2)$
after $(x_1,y_1,z_1)$ in the sequence.  Let $r_3$ be the round in which
$x_3$ became active, and let $r_2$ be the round in which $x_2$ became
active.
Let $s_j = \round_{x_j}[G^i_h](r_j)$, for $j
= 2,3$.
By Lemma~\ref{monotonicLemma},
if $s_2 \ge r'$, then either $x_2$'s knowledge at the
beginning of round $s_2$ is greater than $z_1$'s knowledge
at the end of $r'$, or $z_1 \in F_{x_2}(r')$,
and if $s_2 < r'$, then $z_1
\in F_{x_2}$ before $x_2$ starts operating on $G_i^{h-1}$.  Since $x_2$
sends a message to $z_1$ while operating on $G_i^{h-1}$, it cannot
be the case that $z_1 \in F_{x_2}$ before $x_2$ starts operating
on $G_i^{h-1}$, so it must be the
case that $s_2 \ge r'$ and $x_2$'s knowledge at the beginning of round
$r_2$ is greater than $z_1$'s knowledge at the end of round $r'$.  In
particular, this means that $x_2$ must know that $x_1$ informed $z_1$
about $y_1$ at the beginning of $r_2$.
 
We next show that every
process $x \in G_i^h$ that is active at some round $r$ between
$r'$ and $r_2$ must know that $x_1$ informed $z_1$ about $y_1$ at the
beginning of round $r$.  For suppose not.  Then, by
Lemma~\ref{monotonicLemma}, $z_1$ must have retired by the beginning of
round $r$.  Since, by Lemma~\ref{totalorderLemma}, $x$ is the most
knowledgeable process at the beginning of round $r$, it follows that no
process that is not retired knows that $z_1$ was informed about $y_1$.
Thus, there is no way that $x_2$ could find this out by round $r_2$.
 
It is easy to see that $x_3$ does not know that $z_1$ was informed about
$y_1$ (for if it did, it would not repeat the unit of work $y_1$).
Therefore, $(x_3,y_1,z_2)$ must come after $(x_2,y_2,z_1)$ in the
sequence.  Since $\point_{x_2}[G^i_h](r'') = z_1$, and $z_1$ received
an ordinary message from $x_1$ while operating on $G^i_{h-1}$ at round
$r'$, it follows from
Lemma~\ref{monotonicLemma} that
between rounds $r'$  and $r''$, every process in
$G^i_h$ that is not retired must receive an ordinary message.
In particular, this means
that $x_3$ must receive an ordinary message.
Since all active processes between round $r'$ and $r''$ know that $z_1$
was informed about $y_1$, it follows that $x_3$ must know it too by the
end of round $r''$.  But then $x_3$ would not redo $y_1$, giving us the
desired  contradiction.
\end{proof}

\begin{theorem}
In every execution of Protocol ${\cal C}$,
\begin{itemize}
\item[(a)]
at most $n+2t$ units of real work
are performed,
\item[(b)]
at most
$n+8t \log{t}$ messags are sent,
\item[(c)]
by round
$t(5t+ 2\log t)(n+t) 2^{n+t}$, all processes have retired.
\end{itemize}
\end{theorem}

\begin{proof}
Lemma~\ref{wasteCorollary} implies that the amount of real work
units that are performed and reported to $G_1$
is at most
$|G_0|+|G_1|=n+t$.
In addition, each of the $t$
processes may perform one unit without reporting it (because it
retired immediately afterwards). Summing
the two, (a) follows.
 
For part (b), Lemma~\ref{wasteCorollary} implies that each
$G^i_h, h>0$, performs at most $|G^i_{h-1}|+|G^i_h|$ reported units of works
when operating on $G^i_{h-1}$. (Here a unit is may be either a real work unit
or an `are you alive?' message.)
Let $H = \{(h,i): 1 \le h \le \log t, \
i \equiv 1 \pmod {2^{\log(t) + 1 - h}}\}$.  Notice if we consider
groups of the form $G^i_h$ for $(h,i) \in H$ we count all the groups
exactly once.  The argument above tells us that the total number of
reported units of work is
$$\sum_{(h,i) \in H} (|G^i_{h-1}| + |G^i_h|) \le |G_0| +
3 \sum_{(h,i) \in H} |G^i_h|.$$
The reason for the factor of 3 is that
if $h < \log t$, then $|G^i_h|$ occurs three times
in the left-hand sum: once when considering the work performed by
group $G^i_h$ operating on $G^i_{h-1}$, once when considering the work
performed by $G^i_{h+1}$ when operating on $G^i_h$, and once when
considering the work performed by $G^{i+h}_{h+1}$ when operating
on $G^i_h$.
Clearly, the $|G_0|$ reported units performed on $G_0$
result in one message each, and the
remaining ones result in
two messages each (because then the unit itself is also a message).
So the
number of messages corresponding to reported units of work is at most
$$|G_0| +
6 \sum_{(h,i) \in H} |G^i_h| = n + 6t\log t.$$
 
In addition, the unreported units may result in messages. These
consist both of `are you alive?' messages sent by a process but not reported by
it due to the fact it crashes or terminates immediately afterwards, and of `are
you alive?' messages  that were not reported because the recipient of the `are
you alive?' message responded. Each process in $G^i_h, h>1$ can perform at most
one such unreported unit when operating on $G^i_{h-1}$, and hence each group
$G^i_h, h>1$ performs no more than $|G^i_h|$ such units. In addition, we have to
sum the answers of alive processes in $G^i_{h-1}, h>1$ to `are you alive?'
message sent by $G^i_h$. Again, there are at most $|G^i_h|$ such answers.
Finally, each process $i$ sends messages to the other process in $G^i_{\log{t}}$
just before it starts operating, which together with the answers sums up to a
total of no more than $2t$ messages.
Therefore, the
number of messages corresponding to unreported units of work is at most
$$2t + \sum_{(i,h) \in H, \, h>1} 2|G^i_h| = 2t\log t.$$
 
Summing the messages due to the reported units of work and the messages
due to the unreported units of work,
part (b) follows.
 
Part (c) is immediate from Lemma~\ref{lemma: run time}.
\end{proof}
 
\remove{orli: We remark that we can improve the message
complexity to
$O(t\log t)$ (that is, remove the $n$ term in (b) above)
by informing processes in group $G_1$
after $n/t$ units of work done at level $G_0$, rather than after
every unit of work.  The total work done is still $O(n+t)$; the time
complexity increases to $t(2n+3t + 2\log t)(n+t)2^{n+t}$
because of an increase in
$K$ (the upper bound on the number of rounds, from the
time the currently active process takes over,
that any process needs to wait before first hearing from
the active process).
}

We remark that we can improve the message complexity to
$O(t\log t)$ (that is, remove the $n$ term in (b) above)
by informing processes in group $G_1$
after $n/t$ units of work done at level $G_0$, rather than after
every unit of work.  This does not result in a significant
increase in total work, but it does increase the time complexity.
The increase in time complexity comes from
an increase in $K$ (the upper bound on the number of
rounds, from the time the currently active process takes over,
that any process needs to wait before first hearing from
the active process).  Formally, we have
 
\begin{corollary}
\label{improving the message complexity}
Modifying\notefromO{Made what used to be a remark into a
corollary. The remaining part of the original remark is just after
the corollary.} Protocol \C~by informing processes in group
$G_1$ after $n/t$ units of work done at level $G_0$, rather than
after every unit of work, yields a protocol that sends $O(t
\log t)$ messages, performs $O(n+t)$ work, and terminates
within $t(2n+3t + 2\log t)(n+t)2^{n+t}$ rounds.
\end{corollary}

\section{A Time-Optimal Algorithm}
 
All the algorithms we have
considered so far are inherently sequential: there is only one
process performing work at a time. If processes always have many
(other) tasks that they can do,
then the fact that all but one process is
idle at a given time is not a great problem.  On the other hand,
time is certainly a critical element in many applications.  In
this section, we present an algorithm that aims to achieve maximum
distribution of the work load among the processes.  The algorithm
is time-optimal in the typical case where there are no faults, and
its performance degrades gracefully in the presence of faults.
The basic ideas of this algorithm have been patented \cite{DHS}.
 
The idea of the algorithm is straightforward.
We alternate {\em work phases\/} and {\em agreement phases\/},
until all the correct processes are sure that all the work has
been done. In the first
work phase, process~$j$ performs units
of work $1+jn/t, \ldots, (j+1)n/t$ (we again
assume for simplicity that $n$
is a multiple of $t$) in the first $n/t$ rounds.
Process $j$ starts the first agreement phase by broadcasting a
message to all the other processes saying that it has done its work.
In subsequent rounds,
process $j$ proceeds
much as in
Eventual Byzantine Agreement \cite{DRS}:
It broadcasts its  current {\em view}---what work has been
done, and which processes were alive at the end of the work phase,
from its point of view.
It continues to do so until
(a) the set of processes that are currently
alive, according to $j$'s view, is
the same in two consecutive rounds, or (b)
it receives a message from some process $i$
saying that $i$ is done and containing $i$'s view.
In case (a), it takes as its {\em final view\/} its own
view, while in case (b), it takes as its final view the view
in $i$'s message.
In all cases, it then broadcasts a message
saying it is done, together
with its final view of which processes were alive at the end
of the work phase
and what work remains to be done,
and terminates the phase.
 
Using the by-now standard techniques of \cite{DRS}, we can show that all
the correct processes agree
on their final view at the time when
they terminate the phase, and a correct process
is done
by round $n/t + f+ 2$,
where $f$ is the number of processes that are faulty during the
agreement phase.
Finally, all correct processes terminate at most
one round after the first correct process terminates.
We omit details here.

After process $j$ terminates the first agreement phase, if,
according to its
final view,
$n' > 0$ more work still needs to be done (perhaps because some
process crashed before doing its allocated work)
and $t' \ge t/2$ processes are still correct,
then it starts the second
work
phase.
It performs $n'/t'$ units of work, with the work being divided among the
correct processes according to their id numbers.\footnote{Since
$n'$ may not be divisible by $t'$,
a process might have to do $\lceil n'/t' \rceil$ work.  We ignore
this issue in the discussion, since its impact on complexity is
negligible;
however, the code takes it into account.}
After the work phase, there is an agreement phase, which is just like
the first agreement phase, with one small change.
Whereas in the first agreement phase, if process $j$ did not
hear from process $i$ during some round,
then process $j$ knew $i$ was faulty, in
later agreement phases, since $i$ may be behind $j$ by one step,
$j$ must allow $i$ one round of grace before declaring it faulty.
Similarly, in order to terminate, a process must have two consecutive
rounds {\em after the grace round\/} where,
its view of the set of currently alive processes is the same,
or receive a message from another process saying it is done.
We leave it to the reader to check
that again, at the end of the phase, all correct processes agree
that all the work has been performed, or they agree on their final
view, and that every correct process terminates no more than one
round after the first correct process terminates.
 
We continue in this manner, provided no more than half of the processes
that were correct at the beginning of a phase fail,  until all correct
processes are sure that all work has been done.  If at any phase more
than half the correct processes fail, we revert to another of our
algorithms (for example, Protocol $\A$). We call this
algorithm  protocol \D; the code appears in Figure~\ref{protocol
D code}.\notefromO{Added last two sentences.}
In the code, we use
the function $\grade_S$, where $S$ is a set of nonnegative
integers;  $\grade_S(s) = k$ if there are $k$ elements of $S$
less than $s$.

\begin{figure*}
\begin{tabbing}
XX \= XXXX \= XXXX \= XXXX \= XXXX \= XXXX \= XXXX \kill
 
Main protocol\\ \\
1. \>$S \gets \{1, \ldots, n\}$;\quad\quad \{$S$ is\notefromO{Added
code.} the set of outstanding units of work\}\\
 
2. \>$T \gets \{0, \ldots, t-1\}$;\>\>\>\{$T$ is the set of
processes known to have been correct\\
\>\>\>\>at the end of the previous work phase\}\\

3. \>$\onlytwice \gets 1$;\quad
\{$\onlytwice$ keeps track of whether to allow a grace round\}\\

4. \>{\bf while} $|S| > 0$ {\bf do}\\

5. \>\>$S' \gets \{s \in S: \grade_T(j)\lceil |S|/|T| \rceil \le
    grade_S(s) < (\grade_T(j) + 1)\lceil |S|/|T| \rceil\}$;\\
 
6. \>\>Perform work in $S'$;\\

7. \>\> Wait $\lceil |S|/|T| \rceil - |S'|$ rounds; \quad
\{to make sure all processes
           spend equally long in this phase\}\\
8. \>\>$S \gets S \backslash S'$; \quad\{update outstanding units
of work\}\\

9. \>\> $T' \gets T$;\\
10.\> \>Agree$(\onlytwice)$; \quad \{see code below\}\\
 
11. \> \> {\bf if} $|T'| > 2 |T|$ \quad({\em i.e.} more than half
the processes alive at the end of the previous \\
\>\>\>\> work phase failed by the end of the current work phase)\\
 
12. \>\>{\bf then}\>perform work in $S$ using Protocol $\A$;\\
13. \> \>\> $S \gets \emptyset$;\\
14. \>\>$\onlytwice \gets 0$\\ \\
 
Agree$(\onlytwice)$\\
\\
1.  \>$\done \gets \false$;\\
2. \> $U \gets T$; \{$U$ keeps track of processes not known by $j$ to be
           faulty\}\\
3. \> $T \gets \{j\}$;\\
4.  \>{\bf while} $\neg \done$ {\bf do}\\

5. \>\> $U_j \gets U$; \quad \{save old value of $U$\}\\

6. \> \>Broadcast $(j, S, T,\done)$ to all processes in $U$;\\

7.  \>\>{\bf for} $i \in U_j$ {\bf do} \\

8. \>\>\>{\bf if} received $(i,S_i,T_i,\done_i)$
   {\em and\/} $\done_i = \false$\\
 
9. \>\>\>{\bf then} \>$S\gets S \cap S_i$;\\
10. \>\> \> \>         $T\gets T \cup T_i$;\\
11.\>\>\>{\bf if} received $(i, S_i,T_i,\done_i)$  {\em and\/}
    $\done_i = \true$\\
12. \>\> \>{\bf then} \> $S \gets S_i$;\\
13. \>\> \>\> $T \gets T_i$;\\
14. \>\>\>\> $\done \gets \true$;\\
15. \> \>\> {\bf if} no message received from $i$ {\em and\/}
$\onlytwice \ge 1$\\
16. \>\>\> {\bf then} \>$ U \gets U/\{i\}$;\\
17. \>\>{\bf if} $U = U_j$ and $\onlytwice \ge 1$\\
18. \>\>{\bf then}\> $\done \gets \true$;\\
19. \>\> $\onlytwice \gets \onlytwice + 1$;\\
20. \>Broadcast $(j,S, T,\done)$ to all processes in $U$
\end{tabbing}
\caption{Protool $\D$; Code for Process $j$}
\label{protocol D code}
\end{figure*}

We\notefromO{Maybe we should make a subsection.} now analyze Protocol
$\D$.
The analysis splits into two
cases, depending on whether it is the case that for every phase,
no more than half the processes that
are correct at the beginning of the phase are discovered to have
failed during the phase.

A process $p$ is {\em thought to be correct} at the beginning
of phase $i$ if
$i = 1$ or $i > 1$ and $p$ is in the final view of some
process $p'$ that decided in the phase~$i-1$ agreement protocol.
Note that in the latter case $p$ is in the final view of all
processes that complete the phase~$i-1$ agreement protocol.
\begin{theorem}
\label{protocol D}
In every execution of Protocol \D~in which at most $f$
processes fail,
\begin{enumerate}
\item if for each phase, no more than half the processes
that are
thought to be
correct at the beginning of the phase are discovered to have
failed
by the end of
that phase, then
\begin{itemize}
\item[(a)]
at most $2n$ units of work are performed,
\item[(b)]
at most $(4f+2)t^2$ messages are sent,
\item[(c)]
by round $(f+1)n/t  +4f+2$,
all processes have retired.
\end{itemize}

\item if in some phase more than half the processes that are
thought to be
correct at the beginning of some phase are discovered to have
failed
by the end of the phase, then
\begin{itemize}
\item[(a)]
at most $4n$ units of work are performed,
\item[(b)]
at most $(4f+2)t^2+ 9t \sqrt{t}/(2 \sqrt{2})$ messages are sent,
\item[(c)]
by round $(f+1)n/t+4f+2+nt/2+3t^2/4$, all processes have retired.
\end{itemize}
\end{enumerate}
\end{theorem}
 
\begin{proof}
For part (1), an easy induction on $k$ shows that by the end of
phase $k$, no more than $n/2^k$ units of work remain to be done, and no
more than $n + \cdots + n/2^{k-1}$ units of work have been done.
It follows that at most $2n$ units of work are done altogether.
(We remark that
there is nothing special about the factor ``half'' in our
requirement that we revert
to Protocol $\A$ if more than half the processes that were correct at the
beginning of the phase  are discovered to have failed
 during the phase.  We could have chosen any
factor $\alpha$; a similar proof would show that by the end of phase
$k$, at most $\alpha^k n$ units of work remain to be done, and no more
than $n + \cdots + \alpha^{k-1}n$ units of work have been done, so that
no more than $n/(1-\alpha)$ units of work are done altogether.  However,
it follows from results of \cite{DMY94} that
if we allow an arbitrary fraction of the processes to fail at every
step, and do not revert to Protocol $\A$, it is possible to construct an
execution where $f$ processes fail and
$\Omega(n\log(f)/\log \log(f))$ units of work are done altogether.
Indeed, it follows from the arguments in \cite{DMY94} that this
result is tight; there is a matching upper bound.)
Since each nonfaulty process broadcasts to all the other nonfaulty
processes in each round of an agreement phase, at most $t^2$ messages
are sent in each
such
round.  If $f_k$ is the number of failures discovered
during the $k$th agreement phase, then the first agreement phase lasts
at most $f_1 + 2$ rounds, while
for $k > 1$, the $k$th agreement phase lasts at most $f_k + 3$ rounds,
because of the grace round.
Thus, altogether, the agreement phases last at most
$f+3a-1$ rounds, where $a$ is the number of agreement phases.
Since $a \le f+1$, the agreement phases last at most $4f+2$ rounds,
and at most
$(4f+2)t^2$ messages are sent.
Finally, to compute an upper bound on the total number of rounds, it
remains only to compute how many rounds are required to do the work
(since we know the agreement phases last altogether at most $4f+2$ rounds).
Recall that at the end of phase $k$,
at most $n/2^k$ units of work need to be done.  Since no more than half
the processes fail during any phase, at least $t/2^k$ processes are
nonfaulty.  Thus, at most $(n/2^k)/(t/2^k) = n/t$ rounds are spent during
each
work
phase doing work.
Since there are at most $f+1$ work phases, this gives the required
bound on the total number of rounds.
For part (2), first observe that if we revert to Protocol $\A$ at
the end of phase $k$, then by our earlier observations it is known to the
remaining processes that no more than $n/2^{k-1}$ units of
work remain to be
done, and no more than $(2 - 1/2^{k-1})n$ units of work have been done.
It is also easy to see that at least $t/2$ processes are discovered as faulty.
Moreover, by the bounds in part (1),
at most
$(4f+2)t^2$ messages have been sent and $(f+1)n/t + 4f +2$ rounds have
elapsed.  Now applying Theorem~\ref{protocolATheorem}, we see that at most
$3n/2^{k-1}$ work is performed by protocol~$\A$, no more than $9(t/2)
\sqrt{t/2} = 9t \sqrt{t}/(2\sqrt{2})$ messages are sent, and $nt/2^{k} + 3t^2/4$
rounds are required.
By taking $k = 1$ (the worst case), we get the bounds claimed in the
statement of the theorem.
\end{proof}
 
While the worst-case message complexity of this algorithm
is significantly worse than that
of our other algorithms, the time complexity is better (at least,
if less than half the correct processes fail in each phase).  More
importantly, the situation is particularly good if no process fails or
one process fails.  If no process fails, then $n$
units of work are
done, the algorithm takes $n/t+2$ rounds, and $2t^2$ messages are sent.
If one process
fails, then we leave it to the reader to check that
the algorithm
requires at most $n/t + \lceil n/(t(t-1))\rceil +6$ rounds,
has message complexity
at most $5t^2$,
and at most $n + n/t$ units of work are done.
 
As we mentioned in the introduction, it is easy to modify this
algorithm to deal with a somewhat more realistic setting, where
work is continually coming in to the system.  Essentially, the idea
is to run Eventual Byzantine Agreement periodically (where the length
of the period depends on the size of the work load, and other features
of the system).  We omit further details here.

We can also cut down the message complexity in the case of no failures
to $2(t-1)$, rather than $2t^2$, while still keeping the same work
and time complexity.  Instead of messages being
broadcast during the agreement phase, they are all sent to a central
coordinator, who broadcasts the results.  If there are no failures,
the agreement phase lasts 2 rounds, just as before.
Dealing with failures is somewhat subtle if we do this though,
so we do not analyze this approach carefully here.
 
\section{Application to Byzantine Agreement}
\label{section: Byzantine}
 
One application of our algorithms is to
Byzantine agreement.
A Byzantine agreement protocol
provides a means for $n$ processes, at most $t$ of which may be faulty, to agree
on a value broadcast by a distinguished process called the {\em general},
in such a way that
all nonfaulty processes decide on the same value and, when
the general is nonfaulty, they decide on the value the general
sent.  As in the rest of the paper, we restrict ourselves here to
crash failures.
 
Consider the following Byzantine agreement algorithm. The algorithm
proceeds in two stages: first, the general broadcasts its value to
processes $0$ to $t$; and then, these $t+1$
processes employ one of our sequential algorithms
to perform the ``work'' of informing processes $0$ to $n-1$ about the general's
value.  So,
performing one unit of work here
means sending a message of the form ``the
general's value is $x$.'' To distinguish processes $0, \ldots, t$ from
the others, we refer to them as the {\em senders}.
A few more details are necessary to complete the
description of the algorithm.  First, throughout the
algorithm, each process has a value for the general. Initially, the value is 0.
If a process receives a message informing it about a value for the
general different from
its current value, it adopts the new value. Second, if the
chosen work protocol is $\C$, then we modify it slightly so that each
of its messages contains, in addition to its usual original contents,
the current
value the sender has for the general. Finally, at a predetermined time
by which the underlying work protocol is guaranteed to have terminated,
each process decides on its current value for the general.
 
Observe that processes $0,
\ldots, t$ play two roles in the second stage of the Byzantine agreement
algorithm: they both report the value of the general (as they do work)
and are informed of this value (as work is performed on them
by other senders).

We now prove the correctness of our Byzantine agreement algorithm.
Obviously, if the general is correct, all processes will decide on its
value.
Since at least one of the $t+1$ senders is nonfaulty, it must be
the case that a value is reported by a sender to every nonfaulty
process.
To show agreement, it suffices to show that there is no time at
which two nonfaulty processes that have had a value reported to them
by a sender have different
(current) values.  This, in turn, follows from the
fact that if an active sender $p$ reports a value $v$, and the sender
that was active just before $p$ reported $\bar{v}$, then
at the time $p$ becomes active,
no value was reported to any nonfaulty process.
Assume otherwise. Let $p$ be the first active sender
that violates this claim. Then $p$ reports $v$
for the general and the previous sender reported $\bar{v}$.
Let $q$ be the first sender that was active before $p$ and reported
$\bar{v}$; by construction,
all senders that were active after $q$ but before
$p$ reported $\bar{v}$.
 
By assumption, when $q$ becomes active, no value was reported to any
process that has not yet crashed.  The choice of $q$ guarantees that the
only value that is reported from the time that $q$ becomes active
to the time that
$p$ becomes active is
$\bar{v}$.  It cannot be the case that a value was reported to $p$
during this time, for then
$p$'s value when it becomes active would be $\bar{v}$, not $v$.
\notefromO{Removed since not clear to me why this sentence is
necessary. `It follows that no process can have terminated, for no
process would terminate without reporting a value to $p$.'} In the
case of Protocols $\A$ and $\B$, since work is done in increasing
order of process number, it follows that no value is reported to
any process with a higher number than $p$.  (We remark that it is
important here that a value is {\em not\/} included as part of the
checkpoint messages in Protocols $\A$ and
$\B$.  Since checkpoint messages are broadcast, if a value were included, it is possible that a process
that was active before $p$ crashed while checkpointing to $p$; in this
case, $p$ may not have heard the value $\bar{v}$, and some process with
a higher number than $p$ may have heard it.)
Moreover, the proof of correctness
of Protocols $\A$ and $\B$ shows that all processes with a lower number
than $p$ must have crashed before $p$ became active.  Thus, it
follows that no value was reported to any nonfaulty process at the
time $p$ became active.  In the case of Protocol $\C$, when $p$ becomes
active it is the most knowledgeable nonretired (and hence nonfaulty)
process.  Since for Protocol $\C$ we assume that the checkpointing
messages include the value that was sent, no value can have been sent
to any process that has not crashed.
 
Using Protocol $\C$, we get a Byzantine Agreement protocol
for crash failures that uses $O(n + t\log t)$ messages in the
worst case, thus improving over Bracha's bound of $O(n +
t\sqrt{t})$ \cite{Bracha}.  Using $\A$ or $\B$, we match Bracha's message
complexity, but our protocols are constructive, whereas Bracha's
is not.

\remove{orli: Each of our algorithms can be used to construct an algorithm
for Byzantine agreement along the following lines.
The general sends its value to processes $T= \{0, \dots, t\}$ (note that
at least one of these processes is non-faulty) and then decides on
this value.
The $t+1$ processes then must perform the ``work'' of informing
first the rest of $T$ and then the remaining $n-t-1$ processes of the
value heard from the general.
Thus, the units of work are, in order, informing processes
$1,2, \dots, n$.
At all times a process' current value is the last value it has heard
(0 if it has never heard anything).
When a process becomes active, it informs others of its current value.
At the end of the algorithm it decides on its current value.
 
The proof of correctness of this algorithm varies according to
which of Protocols $\A$, $\B$, and $\C$ is used for performing work.
In the first two cases the proof relies on the fact that processes are informed
about work (more or less) in the same order in which the work was
performed.  In the last case the proof depends on the fact that
the active process is always the most knowledgeable one.
 
We remark that not every algorithm for performing work yields an
algorithm for Byzantine agreement along the lines that
we have described (consider, for example, the trivial algorithm
for performing work, in which
all processes perform all $n$ units of work).}

\section{Conclusions}
In this paper we have formulated the problem of
performing work efficiently in the presence of faults.  We
presented three work-optimal protocols to solve the problem. One
sends
$O(t\sqrt{t})$ messages and takes $O(n+t)$ time, another
requires
$O(t\log{t})$ messages at the cost of significantly greater running
time,
and the third
optimizes on time in the usual case (where there are few failures).  In
particular, in the failure-free case, it takes  $n/t +2$ rounds
and requires $2t^2$ messages.

There are numerous open problems that remain.\notefromO{Please check whether you think all points in these
coclusions are still relevant regarding Yung et al. It seems to
me they still are.}  For example,
it would be interesting to see if message complexity
and running time could be simultaneously optimized. It would also
be interesting to prove a nontrivial lower bound on the message
complexity of work-optimal protocols. Finally, note that by trying
to optimize effort, the sum of work done and messages sent, we
implicitly assumed that one unit of work was equal to one
message.  In practice, we may want to weight messages and work
differently.  As long as the ``weight'' of a message is linearly
related to the weight of a unit of work, then, of course, the
complexity bounds for our algorithms continue to hold.  However,
if we weight things a little differently, then a completely
different set of algorithms might turn out to be optimal. In
general, it would be interesting to explore message/work/time
tradeoffs in this model.

\medskip
\section*{Acknowledgments}
The authors are grateful to Vaughan Pratt for many helpful
conversations, in particular for his help with the
proof of Protocol $\A$.
\remove{orli12: added David Greenberg}
We also thank David Greenberg, Maurice
Herlihy, and Serge Plotkin for their suggestions for improving the presentation
of this work.

\end{document}